*Dielectric spectroscopy of ferroelectric nematic liquid crystals:*
*Measuring the capacitance of insulating interfacial layers*

Noel A. Clark, Xi Chen, Joseph E. Maclennan, Matthew A. Glaser

*Department of Physics and Soft Materials Research Center*
*University of Colorado, Boulder, CO 80309, USA*

*Abstract*

Numerous measurements of the dielectric constant $\varepsilon$ of the recently discovered ferroelectric nematic ($N_F$) liquid crystal (LC) phase report extraordinarily large values of $\varepsilon'$ (up to ~30,000) in the $N_F$ phase. We show that what is in fact being measured in such experiments is the high capacitance of the non-ferroelectric, interfacial, insulating layers of nanoscale thickness that bound the $N_F$ material in typical cells. Polarization reorientation as a linear response to AC electric field-drive renders the $N_F$ layer effectively electrically conductive, exhibiting low resistance that enables the charging of the interfacial capacitors. We analyze the commonly employed parallel-plate cell filled with $N_F$ material of high-polarization $P$, oriented parallel to the plates at zero applied voltage. Minimization of the dominant electrostatic energy renders $P$ spatially uniform and orients it to make the electric field in the $N_F$ as small as possible, a condition under which the voltage applied to the cell appears almost entirely across the high-capacity interfacial layers. This coupling of orientation and charge creates a combined polarization-external capacitance (PCG) Goldstone reorientation mode requiring applied voltages orders of magnitude smaller than that of the $N_F$ layer alone to effectively transport charge across the $N_F$ layer. The $N_F$ layer acts as a low-value resistor and the interfacial capacitors act as reversible energy storage reservoirs, lowering the restoring force (mass) of the PCG mode and producing strong reactive dielectric behavior. Analysis of data from a variety of experiments on ferroelectric liquid crystals (chiral smectics C, bent-core smectics, and the $N_F$ phase) supports the PCG model, showing that deriving dielectric constants from electrical impedance measurements of high-polarization ferroelectric liquid crystals, without properly accounting for the self-screening effects of polarization charge and the capacitive contributions of interfacial layers, can result in overestimation of the $\varepsilon'$ values of the LC by many orders of magnitude.

*Introduction*

The discovery of ferroelectric liquid crystals (FLCs) by Meyer, Liébert, Strzelecki and Keller [1] in 1975 ushered in the era of fluid ferroelectrics and topological ferroelectricity [2]. The novel feature of a macroscopic polarization density that is reorientable and has degrees of freedom absent coupling to any underlying lattice set the stage for a half century of stunning new materials science and applications. Although the polarization of the first ferroelectric liquid crystals was small ($P$ ~ several nC/cm$^2$) [1], the initial experiments demonstrated that these materials were readily reorientable by applied external electric field, manifest as a rotation of the optic axes of the LC and an accompanying change in optical birefringence. It was proposed by Meyer et al. that two kinds of actinic torques could be important in FLCs, those due to external charges (external field torques proportional to $P$), and those due to polarization charges (self-field torques proportional to $P^2$) [1]. The effects of the self-field were demonstrated in early experiments that showed that even in materials with very small $P$, conditions existed under which the self-field could be dominant [3]. In the subsequent proliferation of new smectic FLCs, materials were created with larger and larger polarization, ultimately up to $P$ ~1 µC/cm$^2$ [4]. With increasing polarization magnitudes, self-field interactions were found to produce new modes of fluid FLC behavior, including "block polarization" reorientation [5], "V-shaped" switching [6], and "thresholdless antiferroelectricity" [7], in all of which the polarization charge self-field effectively controls the operative LC structure and field response of electro-optic cells [8].

The field of ferroelectric liquid crystals entered an exciting new chapter with the discovery in 2017 of a ferroelectric nematic (N$_F$) phase in two distinct, new compounds [9,10,11], both exhibiting saturation polarizations $P$ ~6 µC/cm$^2$ [12]. The role of self-fields in ferroelectric nematics has only begun to be explored but such large N$_F$ polarization values clearly push FLC systems firmly into the regime where the self-fields are all-important, with the characteristic length scale balancing elastic and self-field torques decreasing to molecular dimensions [12]. In this paper, we consider the effects of polarization charge and self-field on the electrical behavior of cells having N$_F$ phase material in a gap between flat, conductive plates, focusing on the current vs. voltage, $I(V)$, characteristics of such cells. This type of experiment, with small applied voltages ensuring a linear response, is typically carried out to determine the bulk dielectric constants of liquid crystals, and has been widely applied to smectic FLCs and more recently to ferroelectric nematics. Surprisingly, over the entire history of FLC development, while the electro-optic behavior of such cells has been extensively analyzed, the role of self-fields and polarization charge in dielectric measurements of FLC materials has never been explicitly considered.

[13,14,15,16,17,18,19,20,21]

Liquid crystals with very large polarization values can produce dramatic dielectric effects. For example, numerous publications describing measurements of the dielectric constants ($\varepsilon$) of materials exhibiting the N$_F$ phase [9,11,12] report extraordinarily large values (up to ~ 30,000) of $\varepsilon'$, the real (capacitive) component of $\varepsilon$ in the N$_F$ phase, in many cases claiming $\varepsilon' > 10{,}000$ [9,13-21]. Some bent-core ferroelectric smectic liquid crystals were shown, in earlier experiments, to exhibit similar dielectric behavior [22,23,24], observations which originally stimulated the development of the director-polarization field model we extend here to describe dielectric behavior [4]. As we will show below, properly accounting for the self-fields and polarization charge is critical to understanding the dielectric properties of ferroelectric nematics and other liquid crystal materials with high polarization.

In all of these experiments, high-polarization ($P \gtrsim 1$ μC/cm$^2$) ferroelectric LC samples were introduced into capacitors with known geometries and $\varepsilon = \varepsilon'(\omega) + i\varepsilon''(\omega)$, the dielectric constant of the probed medium, was derived from the measured impedance values using standard relationships between $\varepsilon$ and capacitance; $\varepsilon$ was taken to be uniform and the LC assumed to fill the test capacitor. With these assumptions, the case of widely used sandwich test cells is particularly simple to analyze, in that the electric field within what is effectively a plane-parallel capacitor is uniform, with $E = V/d$, where $V$ is the applied voltage and $d$ the plate spacing. With this condition, and focusing on the real (capacitive) part of $\varepsilon$, the induced displacement field $D = \varepsilon'E$ is also uniform and the resulting induced charge is $Q = DA$, where $A$ is the capacitor plate area, so $Q = DA = \varepsilon'EA = \varepsilon'AV/d = CV$. The dielectric constant is then calculated from $C$ using the familiar relation $\varepsilon' = Cd/A$. A small AC voltage is typically applied as the probe field in these experiments.

While this analysis is straightforward and applicable to a wide variety of purely dielectric materials, it fails if the sample is a fluid ferroelectric with high polarization ($P \gtrsim 0.1$ μC/cm$^2$) [4,5,8], particularly for the geometry employed in the references cited above, in which the applied electric field $E$ in the fluid is substantially normal to the ferroelectric polarization $P$ at low field [25]. In this geometry, field-induced reorientation of $P$ makes $i\varepsilon''(\omega)$ dominantly large, dramatically reducing the effective resistance of the LC layer, from typical values in the megohm to hundred megohm range to values as small, in some cases, as just a few Ohms. This, in turn, renders the interfacial capacitances at the LC/electrode boundaries the dominant series impedance of the cell. With a DC voltage applied to the cell, even tiny, field-induced reorientations of $P$ are sufficient for the self-field to completely cancel the $E$ field in the LC, a negative feedback mechanism which confines the applied field almost entirely to said insulating interfacial layers at the fluid/electrode interfaces: $E$ becomes strongly non-uniform between the electrodes, causing the simple analysis outlined in the previous paragraph to fail.

In this paper, we limit consideration to the dielectric response of FLC cells, under the conditions (*i*) that the polarization is high; (*ii*) that its interfacial layers are insulating; (*iii*) that its interfacial layers are very thin compared with the layer having reorientable polarization; (*iv*) that its temperature is relatively low, in the sense that it is well into the N$_F$ phase; and (*v*) the field-free starting N$_F$ geometry in *Fig. 1* has $\psi \approx 0$, *i.e.*, in the absence of applied electric fields, the director and polarization are oriented parallel to the plates, the alignment strongly preferred by the polarization self-field of the $N_F$. Thus, in dielectric spectroscopy experiments, a small AC electric field, $E_x$, is applied normal to *n* and *P*, so that what is being probed is $\varepsilon_{xx}$, in the notation where the director is along *z*. The component of the dielectric tensor in every equation, figure, and discussion in this paper is $\varepsilon_{xx}$, excluding the cases where an applied DC electric field or magnetic field may be reorienting the director, as noted. These conditions are found in the experimental systems and applicable data being analyzed in *Figs. 3-9*, *C1*. The temperature-independence of key features of the observed spectra confirms that the LC is sufficiently far into the $N_F$ phase that its polarization, *P*, and director, *n* are co-aligned and that the relevant reorientation dynamics of this couple are given by a single Goldstone variable describing the deterministic response of a scalar orientation field to applied electric, elastic, and electrostatic forces. Other phenomena, such as soft-mode polarization variations or amplitude fluctuations are not considered here because they are not necessary to describe well the lower-frequency, lower-temperature dielectric responses and the apparently high dielectric constants claimed in the published analysis of relevant experiments in Refs. [13-21]. However, another major reason for ignoring these effects is that this simplifies the mathematical description of the dielectric response to its essential core, explicitly revealing elements which may not be obvious at first glance, being unique features of high-polarization fluid ferroelectrics, but which are needed to achieve any useful understanding of $N_F$ dielectric behavior.

In order to describe the response of ferroelectric liquid crystals to applied field, we consider a sandwich cell with *P* oriented parallel to the planar cell/electrode plates and normal to *E*. In this geometry, the field applies a torque density $\tau_A = P \times E$ to the polarization field and *P* responds by rotating, a feature unique to fluid ferroelectrics, by an amount given by the collective "Goldstone" reorientation variable $\psi$, as shown in *Fig. 1*. Due to the dominance of the electrostatic energy, *P* is forced to be uniform in the LC [5,8,26] (see *APPENDIX A*), and therefore reorients as a homogeneous block (with $\psi$ constant through the thickness of the cell), generating in the $N_F$ a uniform polarization current flow along *x* which transports polarization charge completely across the $N_F$ layer to where it is confined to the inner surfaces of the (non-polar) insulating (capacitive, $C_i$) layers ubiquitously found at FLC/electrode interfaces. These layers can reduce by orders of magnitude the energy cost of reorientation of *P* relative to that of an isolated $N_F$ layer, intrinsically coupling $\psi$ to capacitive charging and creating what we term a composite polarization-external capacitance Goldstone or "PCG" mode. The reactive energy storage in the charging of the PCG capacitors is the dominant contribution to the reactive restoring force (mass) of the PCG model. Unless the large interfacial capacitance is correctly accounted for in interpreting experimental measurements of the dielectric properties of the $N_F$ phase, the measurements will suggest that the real part of $\varepsilon$ is anomalously large at low frequency. We reiterate that the PCG mode picture emerged, and was uniquely successful, in the effort to understand the electro-optic behavior of high polarization chiral smectic C* and bent-core smectic $AP_F$ materials, and indeed several such examples of this application of the PCG model are included in the discussion below, although our principal focus will be on the $N_F$ phase.

The response of a planar-aligned $N_F$ in a sandwich cell provides a textbook example of the *PCG* mode, which can be summarized quantitively as follows. The applied voltage $V$ provides an electric torque density $\mathcal{T}_A = |\mathbf{P} \times \mathbf{E}| = PV/d$ which activates the Goldstone mode, reorienting the polarization with angular velocity $\omega = \dot{\psi}(t) = PE/\gamma = PV/\gamma d$, where $\gamma$ is the principal nematic rotational viscosity. This rotation generates a current density $\mathbf{J} = \boldsymbol{\omega} \times \mathbf{P}$, resulting in a current $I = AP\dot{\psi}(t) = V[AP^2/d\gamma]$. The current may equivalently be written $I(t) = V(t)/R_{NF}$, where $R_{NF} = d\gamma/AP^2$, which is to say that the $N_F$ bulk behaves electrically as a resistance that decreases with increasing $P$ [4].

The organic material of a ferroelectric nematic is insulating. While there are typically some ionic impurities and resulting conduction in any LC, bulk resistivities of $\rho_{LC} > 10^8$ $\Omega$m are not unusual. A resistor of cross-section $A = 1$ cm$^2$ and length $d = 5$ μm made of a typical liquid crystal in the nematic phase would then have a resistance of $R_{LC} > 5$ M$\Omega$. However, if the sample were an RM734 layer of these dimensions in the $N_F$ phase, then its effective, P-mediated resistance $R_{NF} = d\gamma/AP$ would be $R_{NF} \sim 1\Omega$. This low resistance greatly facilitates current flow through the $N_F$ layer and onto the capacitive insulating layers.

The cell can be viewed electrically as $C_I$ in series with the ($R_{NF}$, $C_{LC}$) parallel combination, as in ***Fig. 1***, where $C_{LC}$ is the bare capacitance of the $N_F$ layer in absence of effects due to $\mathbf{P}$. We can qualitatively compare the behavior of this combination at low frequency in the N phase to that in the $N_F$ phase. In the N phase, $C_{LC} << C_I$ and $R_{NF}$ is large, so that almost the entire applied voltage appears across the LC, over a broad frequency range. Upon transitioning to the $N_F$, $R_{NF}$ becomes very small, effectively shorting out the liquid crystal capacitor $C_{LC}$ and making $C_I$ the dominant impedance in the ($C_I$, $R_{NF}$-$C_{LC}$) series connection at low frequency, so that now the applied voltage appears almost entirely across the insulating layers.

The interfacial layers at the LC/electrode boundaries in $N_F$ cells typically have high resistance and are not ferroelectric. These layers could be alignment layers, oxide or depletion layers, or the surface of the $N_F$ itself where, due to surface pinning, the polarization cannot reorient. The interfacial layers are capacitive in that they can keep the polarization charge on one side and free electron charge on the other separated. They are typically much thinner than the bulk NF layer and therefore have much higher capacitance. They may be regarded as capacitors that are electrically in series with $R_{NF}$, acquiring charge $Q(t)$ at a rate determined by the polarization current passing through the $N_F$, $dQ(t)/dt = I(t)$. It is precisely the capacitance of the interfacial layers that is being measured in dielectric experiments on the $N_F$ phase in this geometry. In a planar-aligned sandwich cell, there are two such layers which have a net capacitance $C_I$. The electrical equivalent circuit of such a cell is a series connection of $C_I$ and $R_{NF}$. The frequency dependence of the response to applied probe fields, including the apparent dielectric relaxation time $\tau_o$, should be understood on this basis ($\tau_o = R_{NF}C_I$). In the steady state, the applied field is entirely screened in the N$_F$ and the entire voltage $V$ applied to the cell appears across the capacitive interfacial layers, up to the voltage $\pm V_{sat}$ where $\psi$ reaches $\pm 90°$.

## Results

*P-C Goldstone mode dielectrics: model and calculation*

The following discussion of the electrical response of N$_F$ cells is based on previously published experimental and theoretical work on block polarization charge screening in liquid crystals and the PCG mode [4,5,8,26,27,28,29,30,31], summarized in *APPENDIX A*, using principally the development of Ref. [4], which deals specifically with the effective conductivity due to Goldstone-mode polarization reorientation in ferroelectric liquid crystals. We consider the sandwich cell geometry shown in *Fig. 1*, with electrodes parallel to the ($y,z$) plane. $\psi$ is the angle between $P$ and $y$, so that $\psi = 0$ in the absence of applied field $E$. In the N$_F$ phase, the nematic director $n$ and $P$ are co-linear.

With a field applied to the LC, $\psi$ becomes time-dependent but remains spatially uniform in high-polarization phases like the N$_F$, i.e., $P$ reorients as a uniform block. This behavior, summarized in detail in *APPENDIX A*, is a result of the potential energy reduction afforded by expulsion of polarization space charge and its associated electric field from the bulk LC to surface layers of thickness

$$\xi_P = \sim 0.15 \text{ nm} \qquad (1)$$

at its boundaries [26]. Here $K$ is a Frank elastic constant and $\varepsilon_{LC}$ is the bare dielectric constant of the LC. The resulting "block polarization" structure, with polarization stabilized kinks (PSKs) confined to the surfaces to match the surface orientational boundary conditions is shown in *APPENDIX A*, in *Fig. A1*. In the N$_F$ phase, due to the high polarization, $\xi_P$ is a sub-molecular-scale quantity and $P$ is a rigidly uniform field extending all the way to the interfaces, with the electrostatic energy cost of maintaining the PSKs overwhelming by orders of magnitude the orientational preferences of typical known nematic surface interactions. As a result, polarization charge transported across the N$_F$ layer is localized, on a sub-molecular scale, to the interfaces where $P$ terminates. The rigidity of $P$ means that the PCG mode is the $q_x = 0$ reorientation mode internal to the N$_F$, and that therefore its description and mathematical treatment by the single scalar variable $\psi$, as sketched in *Fig. 1B*, captures its essential physics. This description must also include the bounding capacitive structures $C_I$, as these strongly affect the electric field in the N$_F$.

Let us now consider the interfacial layers at the LC/electrode boundaries. In the typical LC cell, these layers are much thinner than the LC layer, so we will limit the discussion to this case ($d_I \ll$ *d* in ***Fig. 1***). We also limit the analysis to interfacial layers which are of high enough resistance to be considered to be insulating, so that their effective impedance is just due to the capacitance/area, $c_I$. In a planar parallel cell, there are two such layers, with $d_I$ their combined thickness and $c_I$ their combined capacitance/area. We assume that $\varepsilon_I$ is of order 1, that of typical organic materials.

These capacitive layers, shown in pink in ***Fig. 1***, act to separate the polarization charge at the green/pink interface from the free charge at the pink/yellow interface. Let us consider a positive voltage, *V*, being applied at *t* = 0, as indicated in ***Fig. 1***. At short times, before there is any motion of ***P***, an electric field $E_{NF} = V/d$ appears everywhere in the N$_F$, applying a uniform torque density to ***P***. The polarization field then responds by rotating in the direction that transfers charge to the N$_F$ interfaces, of a sign which acts to reduce the field in the N$_F$. If *V* is less than a saturation voltage $V_{sat}$, then this process ends with $\psi(V)$ such that the interfacial polarization charge $Q(V)$ exactly cancels (*i.e.,* completely screens) the applied field, making $E_{NF} = 0$ and, therefore, the torque on ***P*** also zero. The orientation that ensures that there is no electric field within the N$_F$ is given by the relation

$$\sin[\psi(V)] = c_I V/P = V/V_{sat} \qquad (2a)$$

which couples orientation and applied voltage. This relation holds for |*V*| up to a saturation voltage $V_{sat}$ given by

$$V_{sat} \equiv P/c_I = PA/C_I = PAd_I/\varepsilon_I, \qquad (2b)$$

where for *V* in the range $-V_{sat} < V < V_{sat}$ the range of $\psi(V)$ is $-90° < \psi(V) < 90°$. In the static case, this field-free condition is that which minimizes, with respect to $\psi$, the electrostatic energy of the cell, $U_E(\psi,V)$. Thus, for a given *V*, $U_E(\psi)$ has a parabolic well with a minimum at $\psi(V)$ given by Eq. 2a, established by the electrostatic interactions which, in the case of LCs with high spontaneous polarization *P*, such as in the N$_F$ phase, are dominant over LC elastic and surface forces. The expression in Eq. 2a, which balances the electrostatic torque on $\psi$ due to the applied field with that generated by polarization charge on the N$_F$/insulator interface, can be considered the *master coupling constraint* of the PCG mode (derived explicitly in Eqs. 18,19 below), expressing the coupling between polarization reorientation and applied voltage under static or quasi-static conditions. This is a necessary starting point for calculation of the dielectric response.

We can explore the meaning of $V_{sat}$ by considering the cell with $V = V_{sat}$, and therefore $\psi = 90°$. The field in the N$_F$ due to the polarization charge of magnitude $Q = PA\sin\psi$ which has accumulated at the green/pink interfaces, exactly cancels the electric field in the N$_F$ produced by the free charge of the opposite sign, of magnitude $Q = c_I AV = C_I V$, at the pink/yellow interfaces. This is charge screening effected by the Goldstone reorientation of the polarization field. A steady state has been reached in which there is zero voltage drop across the N$_F$ layer, with the applied voltage *V* distributed across the two interfacial capacitors. $V_{sat}$, the voltage at which the total capacitor charge, $Q = PA$, is the applied voltage at which **P** becomes normal to the plates. Excess voltage beyond $V_{sat}$ is not screened by the polarization (see **Fig. 2**). For an $N_F$ cell with $P$ ~6 $\mu C/cm^2$ and 50 nm-thick insulating layers at the two cell boundaries with $\varepsilon_I = 5$, we estimate $V_{sat}$ ~ 100 V. Therefore, as shown in [12], the $E$-field Freedericksz transition, readily induced by applying a few volts to a sandwich cell in the nematic (N) phase, is strongly suppressed in the $N_F$ (See also Ref. [21]).

For dielectric measurements, the applied voltage is typically only a few volts, so that $V \ll V_{sat}$ and $\psi(V)$ can be taken to be a small angle. In the steady state, when the voltage $V_{NF}$ across the LC is zero, we then have from Eq. 2a

$$\psi(V) \cong V/V_{sat} = c_I V/P = C_I V/PA \qquad (3a)$$

$$Q(V) \cong PA\psi(V) = C_I V. \qquad (3b)$$

At small applied voltages, both the charge $Q(V)$ transferred across the $N_F$ layer and the magnitude of the reorientation $\psi(V)$ are proportional to $V$, meaning that in this DC ($\omega = 0$) limit the cell behaves like a linear, reactive capacitive element. It is important to note, however, that this capacitance is that of the insulating layer, not that of the liquid crystal, which only serves the function of transporting charge through the cell with no voltage drop, like a metal wire.

Under dynamic conditions, the applied voltage $V(t) = V_I(t) + V_{NF}(t)$, where $E_{NF}(t) = V_{NF}(t)/d$, the time-dependent field in the $N_F$, is small but non-zero. The orientation $\psi(E_{NF}(t))$ is the relevant Goldstone variable in the geometry of **Fig. 1** [32,33]. As discussed above, **P** is spatially uniform in the LC, making $E_{NF}$ also spatially uniform. The value of $\psi$ and its dynamics are determined by a balance of torque densities, $\mathcal{T} = PE_{NF}\sin\psi - U\cos^2\psi = \gamma \, d\psi/dt$ (the field + elastic/surface torques are balanced by viscous drag), where $U$ is the elastic/surface torque stabilizing the planar alignment at zero field, and $\gamma$ the rotational viscosity. In the typical electrical excitation of the Goldstone mode in low-polarization FLCs, it is the $U$ term that is reactive, storing potential energy and giving the low frequency $\varepsilon'$ [33,34,35]. In the present PCG case, however, the explicit evaluation presented below of the electrostatic torques maintaining the charge-screening condition in Eqs. 2a and 3a shows that, due to the high polarization, these torques are many (~5) orders of magnitude larger than any possible LC bulk or surface $U$ values. As a result, the coupling of charge transport across the LC layer to the charging of the $C_I$ capacitors makes this by far the dominant potential energy storage mechanism of the PCG mode and therefore the dominant contributor to $\varepsilon'$ at low frequency, leading the uniform field model described above to give apparently giant values of $\varepsilon'$, with no reactive electrical storage in the LC itself.

This quasi-static picture is altered with a time-varying applied voltage $V(t)$, with $\psi$ responding as an overdamped variable to changes in the voltage. With $\psi$ small and spatially uniform in the $N_F$, the torque density acting on the polarization is $\mathcal{T} = PE_{NF} = PV_{NF}/d_{LC} \approx PV_{NF}/d$, where, as before, $V_{NF}$ is the voltage across the LC and $d = d_{LC} + d_I$ is the entire cell gap, where $d_I$ is the total insulating layer thickness. Since $d_I \ll d$, we have $d_{LC} \approx d$. With this notation, the dynamic equation for reorientation of $P$ is

$$d\psi(t)/dt = PE_{NF}/\gamma \approx (P/\gamma d)V_{NF} = (P/\gamma d)[V(t) - \psi(t)\,V_{sat}], \qquad (4)$$

where from Eq. 2a we have $\sin\psi \cong \psi$ for small applied voltages, so in the steady state $V - \psi V_{sat} \cong 0$, and in the dynamic case $V(t) - \psi(t)V_{sat} = V_{NF}(t)$. Since $V(t) = V_I(t) + V_{NF}(t)$, we have $V_{NF}(t) = V(t) - \psi(t)V_{sat} = V(t) - Q(t)/C_I$. This is the small-angle limit of the equation of motion (Eq. 3 in Ref. [4]).

Eq. 4 is recognizable as expressing a form of negative feedback, with $\psi(t)$ always dynamically driven toward the condition $\psi(t) = V(t)/V_{sat}$. Alternatively, the $V(t)$ term can be viewed as a driving flux, and $\psi(t)V_{sat}$ as a flux due to a restoring force or spring constant. In a nematic Goldstone mode, the restoring force is typically due to elasticity or surfaces. In Eq. 4, however, both driving and restoring forces are electrostatic: the steady state condition $V = \psi V_{sat}$ is preferred by very strong electrostatic forces. Eq. 4 is all that is required to describe the dynamics of $\psi(t)$ because the electrostatic torques acting on the director field are orders of magnitude larger than any of the other torques that might be relevant, like the nematic dielectric, diamagnetic, Frank elastic, and Rapini-Papoular torques that are usually important in non-ferroelectric LCs. This is shown explicitly below.

If we define a characteristic time $\tau_o$

$$\tau_o \equiv \gamma d/PV_{sat} = \gamma d C_I/P^2 A, \qquad (5)$$

the dynamic equation for $\psi$ becomes

$$d\psi(t)/dt + [(P/\gamma d)V_{sat}]\psi = d\psi(t)/dt + (1/\tau_o)\psi(t) = (P/\gamma d)V(t) \qquad (6a)$$

or

$$\tau_o\, d\psi(t)/dt + \psi(t) = (C_I/PA)V(t) = V(t)/V_{sat}. \qquad (6b)$$

We note that the time $\tau_o$ for relaxation of $\psi(t)$ decreases as $1/P^2$.

The current flow resulting from small variations of $\psi(t)$ about 0 is $I(t) = PA\, d\psi(t)/dt$, and the charge on the capacitors is $Q(t) = PA\psi(t)$. Using these relationships and Eq. 4, the effective resistance of the LC is found to be

$$R_{NF} = V_{NF}/I = V_{NF}/PA \, d\psi(t)/dt = \gamma d/P^2 A \equiv \rho_{NF} d/A = d/A\sigma_{NF} \qquad (7)$$

where we have introduced the effective resistivity $\rho_{NF} = \gamma/P^2$ and conductivity $\sigma_{NF} = P^2/\gamma$ of the $N_F$. Writing $\tau_o$ in terms of $R_{NF}$, we have

$$\tau_o = \gamma d C_I/P^2 A = (\rho_{NF} d/A)(\varepsilon_I A/d_I) = \rho_{NF}\varepsilon_I d/d_I = R_{NF}C_I, \qquad (8)$$

suggesting that $\tau_o$ may be considered as the time constant of an RC circuit.

The current flow $I(t) = PA \, d\psi(t)/dt$, makes the charge on the capacitors $Q(t) = PA\psi(t)$, enabling Eq. 6 and the PCG stored energy/area $U_{PCG}(Q)$ to be rewritten in terms of charge

$$\tau_o \, dQ(t)/dt + Q(t) = C_I V(t). \qquad (9a)$$

$$U_{PCG}(Q) = \tfrac{1}{2} Q^2/C_I = \tfrac{1}{2} (Q/P)^2 (PV_{sat}). \qquad (9b)$$

Thus, under DC conditions, the $Q/V$ ratio of the cell is just $C_I$, as if all of the applied voltage were across $C_I$, which, as we have seen from the discussion above, is the static case.

In order to calculate dielectric spectra, we let $Q(t) = Q_\omega \exp(i\omega t)$ and $V(t) = V_\omega \exp(i\omega t)$. We then have, from Eq. 9,

$$Q_\omega = C_I V_\omega [1 + i\omega\tau_o]^{-1} = C_I V_\omega \{1/[1 + (\omega\tau_o)^2] - i\omega\tau_o/[1 + (\omega\tau_o)^2]\}. \qquad (10)$$

Returning to expressing $\tau_o = R_{NF}C_I$, we may differentiate Eq. 10, obtaining $I(t) = dQ(t)/dt$, and $I_\omega = i\omega Q_\omega$

$$I_\omega = C_I V_\omega i\omega/[1 + i\omega\tau_o] = V_\omega/[1/i\omega C_I + R_{NF}] = V_\omega/Z_\omega \qquad (11)$$

where

$$Z_\omega = R_{NF} + 1/i\omega C_I \qquad (12)$$

describes the impedance of a series RC circuit. Thus, the I(V) characteristic of the cell is that of a series RC circuit, where the ferroelectric LC provides the resistance $R$, the insulating layer the capacitance C, and $\tau_o$ is the RC time constant. The ratio of the magnitude of the voltage across the $N_F$ layer, $V_{NF\omega}$, to the applied voltage $V_\omega$ across the cell, $|V_{NF\omega}|/|V_\omega| \approx \omega\tau_o$, is small at low frequency where $|Z_{R\omega}| < |Z_{C\omega}|$ and large for $\omega\tau_o > 1$ where $|Z_{R\omega}| > |Z_{C\omega}|$.

Eqs. 10 and 11 represent the dynamic description of a liquid crystal cell with high polarization in the PCG model. An effective experimental approach for characterizing such a cell is simply to measure $Z_\omega$. At low frequency, $\text{Im}Z_\omega$ will determine $C_I$. $\text{Re}Z_\omega$ will be small, peaking at a single-relaxation frequency, $\omega_R = 1/\tau_o$. With $C_I$ and the measured $\tau_o$, we have $R_{NF} = \tau_o/C_I$, and from $R_{NF}$ we get $\rho_{NF} = AR_{NF}/d = \gamma/P^2$, enabling measurement of $\gamma$ for known $P$.

The representation of dielectric cell impedance as a sum of series interfacial and bulk contributions is the phenomenon of "electrode polarization" [36] in the dielectric spectroscopy literature, such that Eq. 12 can be considered a particular case of Eq. 1 in Ref. [36].

*Interpreting dielectric constant data*

The standard approach to measuring $\varepsilon(\omega)$ is rather to convert impedance data into dielectric constant, using a capacitive structure of known geometry, *e.g.*, a specified $d$ and $A$. and measuring its $Z_\omega$. If, for example, the cell simply capacitive, then $Z_\omega = 1/i\omega C_M$ impedance is measured as $Z_\omega = 1/i\omega C_M$ $C_M$, and this measurement is interpreted by assuming that the cell is filled with a homogeneous dielectric, then this gives a measured value of permittivity $\varepsilon_M = C_M d/A$ We call this the homogeneous dielectric (HD) picture.

Starting from Eq. 10, we can calculate from the PCG model what apparent magnitude of the bulk dielectric constant $\varepsilon_A(\omega)$ would be expected from this measuring process when applied to a cell like that of **Fig. 1A**. $\varepsilon_A(\omega)$ is the apparent ratio $D/E$ in the $N_F$ phase, $\varepsilon_A(\omega) = D_{\omega P}/E_\omega$, with $D_{\omega P} = Q_\omega/A$ the contribution to $D$ due to $P$, and $E_\omega = V_\omega/d$. That is, $\varepsilon_A(\omega)$ is the apparent dielectric constant given by actual cell behavior as described by the PCG model, but under the HD assumption of a homogeneous dielectric. We obtain for $\varepsilon_A(\omega)$:

$$\varepsilon_A(\omega) = D_{\omega P}/E_\omega = (Q_\omega/A)(d/V_\omega) = (d/A)[Q_\omega/V_\omega] = (d/A)[I_\omega/i\omega V_\omega] = (d/A)[1/i\omega Z_\omega] \quad (13a).$$

Since $\varepsilon_A(\omega)$ is supposed to be a bulk material parameter and therefore independent of $A$ and $d$, any bulk impedance must therefore be proportional to $(d/A)$:

$$\varepsilon_A(\omega) = (d/A)(1/i\omega Z_\omega) = (dC_I/A)[1 + i\omega\tau_o]^{-1} = \varepsilon_A(0)\,[1 + i\omega\tau_o]^{-1}, \quad (13b)$$

so that
$$\varepsilon_A'(\omega) = \varepsilon_A(0)\{1/[1 + (\omega\tau_o)^2]\}, \quad \varepsilon_A''(\omega) = \varepsilon_A(0)\{\omega\tau_o/[1 + (\omega\tau_o)^2]\}, \quad (14)$$

where at low frequency $\varepsilon_A(\omega \to 0)$ is real and
$$\varepsilon_A(\omega \to 0) = (dC_I/A) = \varepsilon_I(d/d_I), \quad (15a),$$

with the latter equality obtained using $C_I = \varepsilon_I A/d_I$, where $\varepsilon_I$ is the dielectric constant of the insulating layer. Thus interpreting this surface capacitive impedance as coming from bulk dielectric behavior gives an apparent bulk dielectric constant that is thickness dependent, proportional to $d$.

At high frequencies, $\varepsilon_A(\omega > 1/\tau_o)$ is imaginary and

$$\varepsilon_A(\omega > 1/\tau_o) = \varepsilon_A(0)/i\omega\tau_o = (dC_I/A)/i\omega R_{NF}C_I = (d/AR_{NF})/i\omega = \sigma_{NF}/i\omega. \quad (15b)$$

The latter equality shows that the high frequency fall-off of the PCG measured dielectric response is purely *bulk resistive*, the resistance due solely to the PCG mode ($\sigma_{NF} = P^2/\gamma$), the LC filling the cell as a resistive bulk material of impedance $Z_{NF} = R_{NF} = (d/A)(1/\sigma_{NF})$.

So, since $d \gg d_I$, it is the (large) capacitance of the insulating layer, which, if interpreted as coming from the entire cell, gives an (apparently) large $\varepsilon_A(\omega \rightarrow 0)$. There is effectively *no* capacitive contribution from the liquid crystal. This means that if cells from a certain lot with the same cell geometry ($A,d$) and insulating layers ($C_I$) are used in a series of dielectric measurements, then the problematical HD analysis will always yield the same apparent low-frequency, asymptotic $\varepsilon_A'(\omega \rightarrow 0)$ value, independent of the mesogens used, of their temperature, of their degree of planar alignment, of their mixture composition, or of their $\tau_o$ value, as long as they have an $N_F$ phase. This conclusion appears to apply very well to the experimental data from Refs. [19,20] for example, shown in *Figs. 3,4* and discussed further below.

It is interesting to note that the apparent $\varepsilon_A(\omega)$ obtained from Eqs. 10 – 14 of the PCG model (an $\varepsilon_{xx}$) has the same spectrum as that of $\varepsilon_{zz}$ in the single relaxation Born-Debye model for longitudinal polar nematic ordering of a system of free molecular dipoles [37,38]. This is because these models both give the same first-order linear relaxation equations for their respective polarizations [$dP(t)/dt = -(1/\tau)P(t)$], where $\tau$ is in both cases the ratio of a dissipative parameter over a restoring energy, the latter being $k_BT$ in the Born/Debye case and being the electrostatic self-energy $P^2/C_I$ in the PCG case.

*Dielectric constant of the bare ferroelectric nematic*
We can also calculate $\varepsilon_P(\omega) = D_\omega P/E_{\omega LC}$, the contribution of **P** to the dielectric constant of the ferroelectric nematic LC itself, enabling us to explore the claims of observations of "giant" real dielectric constants in the $N_F$ phase. We have $D_{\omega P} = Q_\omega/A$, and from Eq. 4 $E_{\omega LC} = \gamma_\omega/P = i\omega\gamma Q_\omega/P^2A$:

$$\varepsilon_P(\omega) = D_\omega P/E_{\omega LC} = (Q_\omega/A)/(i\omega\gamma Q_\omega/P^2A) = 1/(i\omega\gamma/P^2) = 1/i\omega\rho_{LC} = \varepsilon_{LC}/i\omega\rho_{LC}\varepsilon_{LC} = \varepsilon_{LC}/i\omega R_{LC}C_{LC} \quad (16a).$$

Including the capacitive current through $C_{LC}$ adds the term $C_{LC}E_{\omega LC}d_{LC}/A = \varepsilon_{LC}E_{\omega L}$ to $D$, giving $\varepsilon_{NF}$

$$\varepsilon_{NF}(\omega) = \varepsilon_P(\omega) + \varepsilon_{LC} = \varepsilon_{LC}/i\omega R_{LC}C_{LC} + \varepsilon_{LC} \quad (16b).$$

Here $\varepsilon_{LC}$ is the bare, polarization-free, $N_F$ dielectric constant, and $C_{LC}$ its bare, polarization-free, capacitance (see also *APPENDIX B*). The contribution of **P** to the frequency-dependent dielectric constant of the $N_F$ phase is therefore that of a conducting medium: imaginary with a $1/\omega$ frequency dependence and a characteristic relaxation time that is just the inherent "RC" time constant of the bulk $N_F$ medium. Among ferroelectric liquid crystals, the ferroelectric nematics have the lowest resistivities $\rho_{LC}$ due to their nematic-like viscosities and high polarization. We note again that this $N_F$ conductivity, $\sigma_{LC} = P^2/\gamma$, is non-ionic. This result is independent of $C_I$, and so can be obtained in either the limit $C_I \rightarrow \infty$ or $C_I \rightarrow 0$. In the former, the $N_F$ slab is in direct contact with the electrodes such that $V_{sat} = 0$ and $V_{NF} = V$ in Eq. 4. The same result can be obtained by including $C_{LC}$ in the general treatment of Ref. [4] can be extended to consider the opposite case where $C_I << C_{LC}$ by letting $d_{LC} \rightarrow 0$ with fixed $d$. The general result is again

$$\tau_o \, d\psi(t)/dt + \psi(t) = V(t)/V_{sat}, \qquad (6b')$$

but with $V_{sat} = PA/(C_{LC} + C_I) =$ and $\tau_o = \rho_{LC}d(C_{LC} + C_I)$. This shows explicitly that for a fixed drive voltage the current generated by the relaxation $d\psi/dt$ charges both $C_{LC}$ and $C_I$ acting in parallel, as is clear from Fig. 1C. For $d_I >> d_{LC}$, $C_I << C_{LC}$, the result of Eq. 16b is obtained, and the relaxation time of the Goldstone mode is

$$\tau_{NF} = \rho_{LC}\varepsilon_{LC} = R_{LC}C_{LC}, \qquad (8')$$

the bulk RC time constant of an isolated ferroelectric nematic slab. The low-frequency asymptotic dielectric constant $\varepsilon_A(\omega \rightarrow 0)$ is given by $\varepsilon_A(0) = (dC_I/A) = \varepsilon_I$, as expected since the capacitor is almost entirely filled by a dielectric of permittivity $\varepsilon_I$ in a uniform field.

*Relationship to thresholdless antiferroelectricity*

An important episode in the history of the science of ferroelectric liquid crystals in tilted chiral smectic and polar bent-core liquid crystals is that of "thresholdless antiferroelectricity" and "V-shaped switching". These terms were coined by the Tokyo Institute of Technology group to describe the electro-optic behavior of a family of LCs exhibiting phases having smectic layers with a relatively large polarization, which could be either ferroelectric SmC*, having a uniform polarization (the same in every layer), or antiferroelectric SmC$_A$*, with an alternation of the sign of the polarization from layer to layer [7,39]. Typically, application of an in-layer electric field to an antiferroelectric material could induce a structural transition to the ferroelectric state above a distinct threshold field, as often found in solid state ferroelectrics. In a subclass of these materials, however, such a transition could apparently be achieved without an observable threshold, with the field-induced polarization being linear in the applied field at low field and continuously increasing to ferroelectric saturation at readily achievable fields, a behavior they termed "thresholdless antiferroelectricity". An electro-optic manifestation of this behavior is, at zero applied electric field, extinction of optical transmission through the cell with crossed polarizer and analyzer oriented respectively parallel and normal to the layers. However, in an applied field of either sign, the transmission increases in analog fashion from the $E = 0$ state as the applied voltage is increased, with the principal optic axis continuously approaching the reoriented ferroelectric state, a response termed "V-shaped switching" for its appearance in an oscilloscope trace, and eventually saturating. "Thresholdless antiferroelectricity" was initially attributed to a bulk property of the LCs in question, specifically frustration in the interaction energy of adjacent layers between states of different polarization orientation ("frustroelectricity"), leading to a phase in which the polarization orientations were macroscopically random in the field-free condition.

As it turned out, however, these manifestations of V-shaped switching were not due to such a novel property of an exotic bulk phase. In fact, the phase had been mis-identified as antiferroelectric in the original experiments and was just the basic ferroelectric SmC*, but one having a uniform polarization orientation throughout the cell and manifesting an analog response to applied field, rather than being bistable. The experimental determination (based on optical observations) that the director field must be uniform was initially mysterious but led to the discovery of the "block polarization" state of *P*: uniform orientation as a result of the expulsion of polarization charge from the bulk of the cell, shown in *Fig. A1*. The observed analog response was a result of (*i*) the electrostatic screening characteristics of high polarization in typical LC cell geometries; and (*ii*) the uniquely ferroelectric LC property that the polarization field of a ferroelectric fluid can rotate and not just flip as in solid ferroelectrics [4,5,8,26,27]. As discussed above, this combination leads directly to the PCG model and Eq. 2, which was found to provide a simple, quantitative description of thresholdless (V-shaped) electro-optics, as shown, for example, in *Fig. 2*. The similarity of the application of Eq. 2 to understand "thresholdless antiferroelectricity" and its extension to understand the apparently high dielectric constants of ferroelectric nematics is striking. Both phenomena have been attributed to a bulk property of the ferroelectric LC, but in actuality both are due to the unique properties of ferroelectric LCs with high polarization: the analog response of LCs with high polarization is the electro-optic manifestation of the PCG mode, and the seemingly anomalously high $\varepsilon$ values derived from naïve interpretation of dielectric measurements are an electrical manifestation of the same thing. The PCG model has been confirmed directly in measurements of field-induced optical reorientation, in the SmC* phase as shown in *Fig. 2* and in other examples [4,5,8,26,27], and in the SmAP$_F$ phase [22], as outlined below.

*Development, application, and verification: "smoking guns" supporting the PCG model*
There is a considerable body of theory and experimental observation that critically tests and strongly supports the PCG model. The following examples show that the PCG model, embodied in Eqs. 1 – 15, was developed in the effort to understand the electro-optic behavior of ferroelectric smectic C* phases and of ferroelectric bent-core systems. This work shows that the PCG picture becomes more relevant as the polarization becomes larger, making it especially applicable to the ferroelectric nematics.

This work includes:
- References [4,5,8,26-31] detail the development of the PCG model. In particular, Ref. [4] describes the effective conductivity of the FLC layer, the equation of motion for the $q_x = 0$ PCG

- $\psi(t)$ mode, and the cell impedance shown in Eq. 12. This analysis was not originally extended to the calculation of $\varepsilon_A(\omega)$ and the analysis of dielectric relaxation data, an unfortunate shortcoming that we have rectified here in deriving Eqs. 13-15.
- Electro-optical measurements of $\psi(V)$ as a test of Eq. 2a, showing that $\sin\psi \propto V$ in SmC* [8,5], and SmAP$_F$ cells [22,4,8].
- Tests of Eq. 2b using measurements of the dependence of $V_{sat}$ on $C_I$ in SmC* cells by varying $d_I$, showing that $V_{sat} \propto d_I$ and therefore $V_{sat} \propto C_I^{-1}$ [8];
- Tests of Eq. 2b using measurements on the N$_F$ state of the ferroelectric nematic material DIO, shown in Fig. 5 of Ref. [40]. Here, in response to an applied triangle wave $V(t)$, the polarization reversal current in the N$_F$ is a single peak around $V = 0$, which has a width at its base from $-V_{sat}$ to $+V_{sat}$, which is where $\psi(t)$ is changing. According to Eq. 2b, $Q(t) \propto t$ should have a total width in $V$ of $2V_{sat} = 12$ V and this is what is observed experimentally.
- Measurements on SmAP$_F$ cells showing that $\tau_o \propto d$, as predicted in Eq. 5 [8].
- Measurements on SmC* cells showing that $V_{sat}$ is independent of sample thickness as expected from Eq. 2b [8];
- Measurements on SmAP$_F$ cells showing that $V_{sat} \propto P$, as predicted in Eq. 5 [4].
- The HD assumption that the apparent $\varepsilon_A(\omega)$ obtained in dielectric measurements of high-polarization LCs is a bulk dielectric constant leads to a strange paradox, namely that $\varepsilon_A(\omega)$, which is supposed to be an intrinsic, equilibrium, statistical mechanical property of the ferroelectric phase, increases with cell thickness, in several cases with $\varepsilon_A(\omega) \propto d$ [23,24,41]. To reiterate the discussion of Eq. 15, the apparent linear dependence $\varepsilon_A(\omega) \propto d$ is found if the cells are connected to an instrument for measuring dielectric constants. If the instrument or its operator assumes that $\varepsilon = Cd/A$, and the instrument were to find the same capacitance ($C$) for cells of different thickness, then the vexing result that $\varepsilon_A(\omega) \propto d$ would be obtained. The PCG model, predicting that in the N$_F$ phase the instrument in fact measures $C = C_I$, immediately resolves this conundrum [4].
- Ref [14] reported $\varepsilon_A(\omega)$ data with $\varepsilon_A(\omega) \propto d$, which we now recognize is evidence for the PCG model. This paper included a theoretical effort to account for this surprising observation, based on modeling the internal dynamics of the Goldstone mode. However, the $q_x = 0$ mode and its essential coupling to $C_I$ were left out of this analysis of $\varepsilon_A(\omega)$, leading to the result that $\varepsilon_A(\omega) \propto d$.
- Experimental measurements of the low-frequency, asymptotic $\varepsilon_A'(\omega)$ and $\tau_o$ in planar-aligned N$_F$ cells showing that the measured value of $\varepsilon_A'(\omega)$ is the same for planar cells with $\langle\psi\rangle = 0$ as in cells with $\boldsymbol{P}$ substantially tilted by a magnetic field (with $\langle\psi\rangle = 60°$) [19], consistent with the PCG model prediction that the measured capacitance is $C_I$ in all cases.
- Measurements on SmAP$_F$ cells with interfacial layers which have a resistance $R_I$ in parallel with $C_I$. For $\omega < (R_I C_I)^{-1}$ these layers are resistive, so the overall equivalent circuit has two resistors in series. If $R_I$ is large compared with $R_{NF}$ then the measured $\varepsilon_A(\omega) \propto (i\omega R_I)^{-1}$ [17].
- Analysis in Ref. [4] of the response of a high-polarization SmAP$_F$ cell to an applied triangle wave providing confirmation of the overall block polarization picture. When driven by a triangle-wave voltage, the capacitive charging current is I = dQ(*t*)/dt ∝ d*V*(*t*)/dt ∝ constant. For $-V_{sat} < V(t) < V_{sat}$, the capacitive interfaces $C_I$ undergo constant-current, low-resistance charging, with the current suddenly dropping if the applied voltage exceeds $V_{sat}$, giving, for a general triangular driving voltage, a flat-topped polarization current peak centered about $V = 0$. A particularly striking observation is that ion current appears only when $V > V_{sat}$, i.e., when, as in *Fig 2*, $V_{NF}$ becomes a substantial fraction of $V$, direct evidence that $V_{NF} << V_{sat}$ for $V < V_{sat}$ [4].

- Measurement and analysis of $\tau_o$ for DIO at $T = 60°C$ using the data in Ref. [**19**]. From the peak in $\varepsilon''(\omega)$ in *Fig. 3A* for $B = 0$, we have $f_o \approx 7$ kHz and therefore $\tau_o \approx 2 \times 10^{-5}$ sec. This value can be estimated quite well using $\tau_o = R_{NF}C_I = \rho_{NF}\varepsilon_I d/d_I = (\gamma/P^2)\varepsilon_I d/d_I$ from Eq. 8 and assuming $\gamma = 0.25$ Pa-s [42], $P \approx 6 \times 10^{-2}$ C/m$^2$ [42], $d = 20$ μm [19], $\varepsilon_I \sim 5\varepsilon_o$, and $d_I \sim 4$ nm, which gives $\tau_o \approx 1.4 \times 10^{-5}$ sec, in good agreement with experiment.

- The disappearance in the ferroelectric phase of the "ion feature" seen at low frequency in the phase above the ferroelectric phase in temperature, in scans of both $N_F$ and SmAP$_F$ materials [20,15,23,43]. This effect, which is illustrated in *Figs. 4,5,7,8*, is discussed in the Section on *Ion effects*.

- One of the most striking changes in LC textural phenomena due to the transition to the ferroelectric nematic state is the apparent loss of the Fredericksz transition: in typical planar-aligned cells, applied electric fields which readily produce the Freedericksz transition in the nematic phase have no observable effect on the $N_F$, the threshold field apparently having become orders of magnitude large in the $N_F$ phase. This behavior is opposite that initially expected for the $N_F$ phase [9], in which initial study showed an apparently very large dielectric anisotropy [9], which should make for a correspondingly low Freedericksz threshold field. Observation of suppression of the dielectrically- or magnetically-induced planar-to-homeotropic Freedericksz transition in the $N_F$ phase. This is discussed in the Section on *Orientation by surface treatment and applied field*, *Fig. 3*, and Refs. [12,19,21].

- Measurement of $\varepsilon_A'(\omega \rightarrow 0)$ of an $N_F$ phase in a sandwich cell with bare gold electrodes gives $\varepsilon_A'(\omega \rightarrow 0) \approx 12000$, but a similar cell with thicker polymer layers on the gold gives $\varepsilon_A'(\omega \rightarrow 0) \approx 1300$ [44]. The latter case has a nanoscale thick insulating layer at the gold/$N_F$ interface, making $C_I$ large, but the polymer layer is ~10x thicker and $C_I \sim 1/d_I$ correspondingly smaller.

- *APPENDIX C* analyzes recent broadband measurements of the dielectric constants of the $N_F$ phase of RM734 and DIO which provide the first full dielectric spectra of ferroelectric nematics as a function of frequency, sample thickness and temperature. These data provide strong evidence for the PCG model and eliminate in a model-free fashion any possibility of large bulk $\varepsilon$ values in the $N_F$.

*Characteristic frequency dependence of the PCG mode: Comparison with literature measurements*
The PCG dielectric model can be compared to the experimental measurements of the frequency and temperature dependence of the apparent $\varepsilon_A'(\omega)$ and/or $\varepsilon_A''(\omega)$ presented in Refs. [9, 13-24].

These data were originally analyzed using the HD model (Eqs. 13-15), with the authors determining the dielectric properties assuming that the cells were filled with homogeneous dielectrics of permittivity $\varepsilon_A(\omega)$, leading to claims of having measured giant $\varepsilon_A'(0)$ values. We find, in contrast, that these results are all consistent with the PCG model and that there are, in fact, no large capacitive dielectric anomalies in the $N_F$ phase.

The experimental data in Refs. [9, 13-24] are not presented as raw $Z_\omega$ values from which $R_{NF}$ and $C_I$ could be calculated, but in the PCG model, since $\varepsilon_A(\omega) \propto Q_\omega$, the frequency dependence of the apparent $\varepsilon_A(\omega)$ is still that given by Eq. 10, from which $\tau_o$ can be obtained and the predicted, single-relaxation relationship of the PCG model between the real and imaginary parts of $Q_\omega$ tested. Fits of the experimental frequency scans using Eq. 10 are shown in *Figs. 3-9*, with the temperature of each fitted scan selected so that the sample is in the ferroelectric $N_F$ or SmAP$_F$ phase. These materials all exhibit characteristic relaxation behavior, with a single relaxation time $\tau_o$ in the .01 – 100 msec range that can be simultaneously matched to the real and imaginary features of $Q_\omega$ by scaling of the two axes. The PCG model clearly fits the experimental data well. In principle, $Z_\omega$ could be extracted from these data and used to obtain $C_I$ and $\tau_o$, and then Eq. 5 used to get a measurement of $\rho_{NF} = AR_{NF}/d = \gamma/P^2$, and thence obtain $\gamma$ for known $P$.

*Orientation by surface treatment and applied field*
Dielectric constant measurements typically employ applied fields that are sufficiently small that they do not substantially perturb the director field alignment promoted by the cell surfaces and which establish the orientation of molecules nearby by local interaction. In an FLC, the surfaces also serve to terminate the polarization: ***P*** = 0 outside the liquid crystal, so the component of ***P*** normal to the surface deposits polarization charge, filling space with macroscopic electric field. This leads generally to a preference when the LC is between parallel plates for the formation of macroscopic patterns of polygonal domains with uniform ***P*** in which ***P*** is parallel to the sample surfaces and normal to interior domain boundaries [45], a form of block polarization charge-screening leading to textural motifs that minimize the macroscopic electric field. Analogous domain structures are also familiar from ferromagnets [46]. The simplest manifestation of such a texture in the $N_F$ phase occurs in planar-aligned sandwich cells, with ***P*** parallel to the plates. This alignment is observed to be definitively preferred in the $N_F$, as reported in the prototype ferroelectric nematic RM734 [10,12]: in the N phase, a director orientation normal to the surfaces is readily induced by a Freedericksz transition but such a transition is strongly suppressed upon cooling into the $N_F$ [12], which we show in what follows to be a block polarization screening effect. Polarization screening may play an additional important role in $N_F$ dielectric behavior beyond facilitating the PCG mode, in that it also participates in establishing the director and polarization alignment in the cell.

*Electrostatic stabilization of planar alignment* – The role of block polarization screening in establishing planar alignment can be quantitatively assessed theoretically with the aid of ***Fig. 1B***, which sketches a state of intermediate orientation with non-zero $\psi$ in which the charge on the capacitive surface layers is $Q(\psi) = PA\sin\psi$. This state is obtained from the starting, uncharged $\psi = 0$ state by uniformly applying, via an external mechanical agent for example, a quasi-statically increasing torque density $\mathcal{T}(\psi)$ to the N$_F$ *P-n* couple until the desired $\psi$ is reached, a process carried out at constant *V*. We begin with $V = 0$, *i.e.*, with the overall cell shorted. In this case, $V_I = -V_{NF}$ and $E_{NF}d_{LC} = -E_I d_I$, so that $|E_{NF}| \ll |E_I|$, and $E_{NF}$ is small field in the N$_F$ that provides the restoring torque. Increasing $\psi$ from zero does work charging the capacitors. The ratio, *R*, of energy stored in the interface layers to that stored in the LC is given by $R \approx d_I E_I^2 / d_{LC} E_{NF}^2 = d_{LC}/d_I \gg 1$, again assuming that $\varepsilon_I$, $\varepsilon_{LC}$ are of order 1. Thus the electrostatic energy cost of changing $\psi$ is taken as that on $C_I$ $U(Q(\psi), V = 0) = Q^2/2C_I$. For a given $\psi$, the restoring torque/volume is given by

$$\mathcal{T}_R(\psi)Ad = -dU/d\psi = -d(Q^2/2C_I)/d\psi = -(Q/C_I)dQ/d\psi = -(PA)^2\sin 2\psi/2C_I = -U_{max}\sin 2\psi \quad (17a),$$

or
$$\mathcal{T}_R(\psi) = -(PV_{sat}/2d)\sin 2\psi \quad (17b),$$

where $U_{max} = (PA)^2/2C_I = (PAV_{sat}/2)$ is the total energy deposited in the surface layers at $\psi = 90°$. Note that the energy, $U_{max}$, deposited in this case where the cell is shorted, is the smallest energy required to reorient $\psi$ by 90° in such a cell. In contrast, in a cell having an interfacial insulating layer that is much thicker than the LC layer, which approaches being equivalent to an open-circuited cell, we would have $U_{max} \approx (PA)^2/2C_{LC}$, larger by the ratio $d_{LC}/d_I \gg 1$. In this limit, the energy storage is the largest achievable given the available charge/area, *P*.

*Electric field-induced reorientation of the N$_F$: derivation of the master coupling constraint* – Let us consider an externally driven torque density generated electrically. Applying a voltage *V* to the cell gives an electric field on the LC of $E_{NF} = V/d$ and an applied torque density $\mathcal{T}_A = |\mathbf{P}\times\mathbf{E}| = (PV/d)\cos\psi$ coupling to the polarization field. The effective energy density $U(Q(\psi),V)$ and the torque density $\mathcal{T}_N(\psi)$ controlling the variable $\psi$ will then be, respectively,

$$U(Q(\psi),V) = Q^2/2C_I - \mathbf{P}\cdot\mathbf{E} = (PA\sin\psi)^2/2C_I - (PV/d)\sin\psi. \quad (18a)$$

$$\mathcal{T}_N(\psi) = \mathcal{T}_A(\psi) - \mathcal{T}_R(\psi) = (PV/d)\cos\psi - (PV_{sat}/2d)\sin 2\psi = (P/d)\cos\psi[V - V_{sat}\sin\psi]. \quad (18b)$$

The fact that $\mathcal{T}_A$ and $\mathcal{T}_R$ have different dependence on $\psi$ means that $\psi$ is electrostatically driven toward a particular $\psi(V)$ at each *V*, defining the existence of the master coupling constraint and determining its functional dependence, in this case $\sin\psi(V) = V/V_{sat}$, which gives the $\psi$ value that minimizes the electrostatic energy and, as noted in Eq. 2a, gives $E_{NF} = 0$. This relationship has been tested extensively by observing SmC* director reorientation under a triangle-wave driving voltage (see, for example, ***Fig. 2***).

The magnitude of the electrostatic energy enforcing the $\sin\psi(V) = V/V_{sat}$ driving imposed by the master coupling constraint can be obtained by calculating $d\mathcal{T}_N(\psi)/d\psi$, which is given by

$$d\mathcal{T}_N(\psi)/d\psi|_{\sin\psi(V) = V/V_{sat}} = (P/d)[V^2 - V_{sat}^2]/V_{sat} = -(PV_{sat}/d)\cos^2\psi. \quad (19a)$$

Thus, the torque keeping the polarization at $\psi = \sin^{-1}(V/V_{sat})$ is largest for small $\psi$ and $V$. At $V = 0$, this torque tends to keep the polarization at $\psi = 0$, acting as an effective surface anchoring agent favoring planar alignment, with a ferroelectric Rapini-Papoular -like energy/area given by $W_F$

$$W_F \equiv d[d\mathcal{T}_N(\psi)/d\psi] = -PV_{sat}\cos^2\psi \approx -w_F + \tfrac{1}{2}w_F\psi^2, \qquad (19b)$$

where $w_F = PV_{sat}$, and the final approximation is for $\psi$ small. For a typical $N_F$ polarization of $\sim 6 \times 10^{-2}$ C/m², with very thin insulating layers with $d_I = 2$ nm and $\varepsilon_I = 5$ at the two cell boundaries, we estimate $V_{sat} \sim 6$ V. It seems unlikely that $V_{sat}$ values for such $N_F$ materials would ever be much less that 1 V, so we estimate $w_F = PV_{sat} = P^2/c_I > 6 \times 10^{-2}$ J/m² as their smallest effective electrostatic anchoring coefficient. The corresponding $c_I$ range is $c_I < 0.1$ F/m². This estimate can be converted into one for the largest electrostatically controlled $\varepsilon_A(0)$ by using Eqs. 19b and 15 to write $w_F = PV_{sat} = P^2/c_I = [P^2/\varepsilon_A(0)]d_{LC} > 6 \times 10^{-2}$ J/m². For $d_{LC} = 10$ μm, we find $\varepsilon_A(0) < 10^5$.

For commonly used polymer alignment layers, one has $d_I \gtrsim 20$ nm, $V_{sat} \gtrsim 100$ V, and values of $w_F$ in the 1 J/m² range can be expected. In contrast, the Rapini-Papoular coefficients for typical surface alignment treatments in nematic cells are in the range $10^{-4} > w > 10^{-7}$ J/m² [47], much smaller than that in the $N_F$ phase: the intrinsic electrostatic preference for planar alignment dominates any non-planar alignment preferred by rubbing or other common surface treatments by many orders of magnitude. An example of this can be found in Figs. S1,S2 of Ref. [21], where homeotropic surface treatment was found to produce substantially planar cells. More generally in the $N_F$ phase, where $\xi_P$ is of atomic scale, electrostatic interactions dominate the internal structure on all larger length scales and $\psi(r,t)$ is determined by minimizing the electrostatic energy. In the case of dynamic reorientation, viscous forces also come into play. The previous paragraph shows that electrostatic interactions are likely to control the restoring force of the PCG mode.

*Quadrupolar field-induced reorientation of the $N_F$* – We also consider the response of $N_F$ cells to an externally generated, quadrupolar torque density applied to the liquid crystal either magnetically or dielectrically with an electric field of sufficiently high frequency that only the background positive dielectric anisotropy of the $N_F$ is relevant. In the magnetic case, we have a net torque density

$$\mathcal{T}_N = \mathcal{T}_A - \mathcal{T}_R = (\Delta\mu H^2/2)\sin 2\psi - (PV_{sat}/2d)\sin 2\psi = [\Delta\mu H^2/2 - PV_{sat}/2d]\sin 2\psi. \quad (20)$$

We see that, at the level of the approximations employed thus far, $\mathcal{T}_A$ and $\mathcal{T}_R$ in this case have the same dependence on $\psi$, so that all values of $\psi$ become unstable simultaneously above the threshold field where $\Delta\mu H_c^2/2 = PV_{sat}/2d$. In this situation, secondary effects, such as remnant deformations of $P(r)$, will determine whether the induced tilt of $P$ grows continuously from $\psi = 0$ as the field is increased from zero, or jump, as in a first-order transition, to finite $\psi$ with a sufficiently large field. We do not pursue this question here, but the measurements of Nishikawa *et al.* [19] show that with an applied magnetic field, $\psi$ increases smoothly from $\psi = 0$ with increasing field, indicating that in their experiments $\mathcal{T}_R$ increases relative to the magnetic $\mathcal{T}_A$ with increasing $\psi$. Even without understanding of this detail, we can make a semi-quantitative estimate of the quadrupolar field necessary to induce homeotropic alignment of the N$_F$, by comparing the available torque density with $\mathcal{T}_R$ in Eq. 17, as follows.

The quadrupolar transition to the homeotropic state begins at the Freedericksz threshold, where the torque density of the applied field, $\mathcal{T}_A$, becomes comparable to the energy density of the fundamental director distortion mode in the cell of thickness, $d$, given by $\mathcal{T}_{Freed} \sim \Delta\mu H^2 \sim \Delta\varepsilon E^2 \sim K_S(\pi/d)^2$, which, with splay elastic constant $K_S \sim 5$ pN and cell thickness 10 μm, corresponds to $\mathcal{T}_{Freed} \sim 0.5$ J/m$^3$. Molecules are substantially tilted (to $\langle\psi\rangle > 70°$) in the N phase for $\mathcal{T}_A$ several times $\mathcal{T}_{Freed}$. In the N$_F$ phase, the torque density required for similar reorientation is, from Eq. 17b, $\mathcal{T} \sim \mathcal{T}_R \sim PV_{sat}/2d \sim 3 \times 10^5$ J/m$^3$, $\sim 10^5$ times larger than $\mathcal{T}_{Freed}$ for the Frank elastically-stabilized N phase. This huge torque density requirement accounts for the suppression of the Freedericksz transition upon passing from the N to the N$_F$ phase in planar-aligned RM734 cells reported in the *SI Appendix* of Ref. [12]. A similar result is reported for DIO in Ref. [21]: "In the N$_F$ phase the splay elastic constant cannot be determined as no influence of electric field on dielectric constant was found, up to the high voltage that destroys sample alignment" [21].

Magnetic reorientation of *P* in the N$_F$ phase has been demonstrated by dielectric spectroscopy [19]. Here $\varepsilon_A(\omega)$ measurements were carried out on $d = 20$ μm plane-parallel cells with ITO electrodes treated with octadecyltriethoxysilane. Although this treatment typically gives homeotropic alignment of nematics, in these cells the N$_F$ aligned with *P* parallel to the plates, in agreement with the observations of Caimia *et al.* [48]. In Supplementary Fig. 7 of Ref. [19], reproduced in *Fig. 3A* below, the aligning magnetic field *B* was applied both parallel and perpendicular to the plates. Model spectra calculated using Eq. 10 of the PCG model are overlaid on these data (magenta symbols), providing an excellent match of the spectral features, with the two PCG parameters adjusted here to fit the $B = 9$ T data. With the field applied in-plane in the range $0 < B < 9$ T, even a very large *B* produces very little change of the dielectric response *Fig. 3A* (right), consistent with the field simply reinforcing the pre-existing alignment parallel to the plates, with the probe field *E* everywhere normal to *P*. Interpreted in the context of the RC circuit analog of the PCG model, this insensitivity to *B* means that the measured low-frequency capacitance ($C_I$ in the PCG model), proportional both to $\varepsilon_A'(0)$ and to the RC time constant $\tau_o$, does not change with field.

With *B* applied normal to the cell plates, the case shown in *Fig. 3A*, $\tau_o$ increases with increasing *B*, as seen in *Fig. 3C*. In order to understand this result, we assume that the applied magnetic torque reorients $\psi$ to a new average value $\langle\psi\rangle$, at which orientation the field-induced torque on *P* becomes $P_xE_{NFy} = PE_{NF}\cos\langle\psi\rangle$, and the current becomes $I(t) = P\cos\langle\psi\rangle A$, reducing the LC conductivity to $\sigma = (P\cos\langle\psi\rangle)^2/\gamma$ and thereby, according to Eq. 7, increasing $\tau_o$. Assuming this dependence of $\langle\psi\rangle$ on $\tau_o$, we can extract the variation of $\langle\psi\rangle$ with $B$, which we show in *Fig. 3C*, with $\langle\psi\rangle$ reaching ~45° for $B$ = 9 T. For comparison, the splay-bend Freedericksz transition threshold for magnetic alignment in a typical nematic such as 5CB is $B_{Freed}$ ~ 0.1 T, with $\langle\psi\rangle$ ~ 45° attained at $B$ ~ 0.2 T. Thus, comparable reorientation of the director normal to the bounding plates in the N and the N$_F$ phases requires a $\mathcal{T}_A$ that is ~$(9/0.2)^2 \approx 2000$ times larger in the N$_F$ phase in these experiments.

In the HD model, such a field dependence of $\tau_o$ would be interpreted as a consequence of quadrupolar dielectric anisotropy of the N$_F$ phase. But this approach is complicated by the fact that even where the $B$-field is applied normal to the bounding plates, where the observed $\tau_o(B)$ indicates that $\langle\psi\rangle$ is changing, there is essentially no dependence on $B$ of the low-frequency asymptote of $\varepsilon_A'(\omega)$, which would certainly be expected to exhibit some anisotropy if $\langle\psi\rangle$ is changing.

Dielectrically related behavior is reported by the same authors in Supplementary Fig. 17 of Ref [20], reproduced here in *Fig. 4*. While the $\tau_o$ values in a series of mixtures of varying chemical composition studied over their entire N$_F$ phase range of temperatures change by factors up to ~500, essentially no dependence of the low-frequency asymptote of $\varepsilon_A'(\omega)$ was observed in these experiments. Therefore, $\varepsilon_A'(0)$ appears to be independent of the changing properties of the N$_F$ materials investigated. These observations are readily understood using the PCG model, providing strong direct evidence supporting this model: the capacitance measurement from which $\varepsilon_A'(0)$ is being calculated, is, in the PCG model, only probing $C_I$. This happens in Ref. [19] because $C_I$, a property of their carefully prepared silane insulating interfacial layers, is apparently independent of $\psi$, and in Ref. [20] because the same kind of cell, with the same carefully prepared silane insulating interfacial layers, is used for all of the measurements reported there. Since $\tau_o = R_{NF}C_I$, the variation of $\tau_o$ can be attributed to changes in $R_{NF}$, *i.e.*, in $P$ and $\gamma$.

Given the information provided about the cells used, we can make a quantitative estimate of the value of $\varepsilon_A'(0)$ to be expected in the experiments of Ref [20], using Eq. 15 of the PCG model. The insulating interface layer is OTE silane, which, for this purpose, we assume to be a close-packed monolayer of thickness $d_I \approx 2.5$ nm (the extended all-trans OTE length, 20 x 0.126 nm) with the chains normal to the surface plane. We take the OTE layer dielectric constant as $\varepsilon_I \approx 2$, which is the typical DC ($\omega$ = 0) alkane liquid value near room temperature [49]. With the given cell thickness of $d$ = 13 µm [**20**], we find $\varepsilon_A'(0) = \varepsilon_I(d/d_I) = 2(13,000/2.5) \approx 10,400$, which compares well with the measured value of 13,000. This result indicates that if the chains are not tilted, there is, on the nanometer scale, very little in the way of insulating layers in addition to the OTE. Tilting is common in self-assembled monolayers of alkyl chains and would result in smaller $d_I$. If the OTE chains were in fact tilted, there would be contributions to $d_I$ on the 1 nm-scale coming from the ITO surface or some kind of immobile boundary layers in the N$_F$, which appear to be present on pure gold surfaces [44]

*Polarization charge screening effects in the antiferroelectric phase*

The electric field-induced transition from the antiferroelectric SmZ$_A$ phase to the N$_F$ is similarly affected by this phenomenon, as illustrated in the current response of the ferroelectric nematic material DIO plotted in Fig. 5 of Ref. [40]. In order to observe this transition in a cell, the condition $V > V_{sat}$, where $V_{sat}$ is calculated for the ferroelectric state, must be satisfied for $E_{NF}$ to be large enough to induce and stabilize the field-induced ferroelectric state. This experiment was carried out using a cell with small $V_{sat}$: DIO was filled into a thick ($d = 100$ μm) sandwich cell with very thin insulating layers (bare ITO electrodes), generating planar alignment of the polarization in the SmZ$_A$ phase. Reorienting the polarization to point along the direction normal to the cell plates was achieved for $|V| > V_{sat}$ ~12 V in this cell.

*Ion effects and the paraelectric phase*

All of the sets of $\varepsilon_A'(\omega)$ scans in *Figs. 4,5,7,8* exhibit "ion features" at low frequency in the phase above the N$_F$ in temperature, highlighted by the green circles on sample sets. In *Fig. 8*, this feature is a well-defined ion peak, which can be seen in $\varepsilon_A''(\omega)$ at $\omega_{ion}$ in the range 10 Hz < $\omega_{ion}/2\pi$ < 100 Hz. In the *Fig. 4,5,7* data sets, where $\omega_{ion}$ is smaller than the lowest frequency shown on the plot, the ion feature appears as an $\varepsilon_A'(\omega) \propto 1/\omega$ tail at low frequency. There are several common behaviors in the temperature dependence of the ion feature in all of the data sets which can be understood using the PCG model, and provide further evidence of the model's validity:

(*i*) The ion features in *Figs. 4,5,7,8* disappear abruptly on cooling into the N$_F$ phase, as is shown dramatically in *Fig. 8*. Typical dielectric measurements in the N$_F$ phase, such as those shown in *Figs. 3-8*, show $\varepsilon_A'(\omega \to 0)$ values that are independent of $\omega$ over decades of decreasing frequency in the regime where $\omega\tau_o < 1$. This is the ideal PCG-mode response, dominated by polarization current in the N$_F$ liquid crystal layer feeding charge to ideal, insulating, capacitive interfacial layers. The ion features in the higher temperature phases, the result of current flow provided by ionic charge carriers in the LC layer, disappear on entering the N$_F$ phase because in the N$_F$ the ion current is "shorted out" by the polarization current, *i.e.*, $R_{NF}$ associated with reorientation of the spontaneous polarization is much lower than that due to the ions and is in parallel with the ion current, so the appearance of $R_{NF}$ dramatically reduces $E_{NF}$ and effectively eliminates the ion current.

(*ii*) Common features of the temperature dependence of $\varepsilon_A'(\omega)$ in *Figs. 4* (80/20 and 70/30 mixtures) and *Fig 8* are highlighted inside the green and cyan circles in *Fig. 8*. The cyan circle indicates a frequency range where $\varepsilon_A'(\omega)$ is independent of $\omega$ and increasing on cooling. The corresponding $\varepsilon_A''(\omega)$ (orange circle) is suggestive of "soft mode" behavior of the paraelectric phase (the N phase in *Fig. 4* and the smectic A in *Fig. 8*). A PCG-like model to account for $\varepsilon_A(\omega)$ in this $\omega$ range would have the soft mode acting as a low-resistance current source supplying charge to a surface capacitive layer that is becoming thinner with decreasing $T$. In this regime, $\omega$ is high enough that the ion motion does not contribute to the current but as $\omega$ is decreased into the range encompassed by the green circle, $\varepsilon_A'(\omega)$ increases and the ions come into play. Below this ionic relaxation frequency, $\varepsilon_A'(\omega)$ increases to its ultimate frequency-independent $N_F$ value. This shows that the ionic current in the paraelectric phase can transport charge onto the $N_F$ surface layers as effectively as the polarization current does in the $N_F$, charging up the same insulating capacitive layers in both cases. The lower $\varepsilon_A'(\omega)$ values in the cyan circle mean that the soft-mode transport leaves a thicker charge-free layer at the surface, which layer becomes continuously thinner with decreasing $T$, ultimately approaching its condition in the $N_F$.

(*iii*) The temperature dependence of $\varepsilon_A''(\omega)$ in *Fig. 8* in the orange circle indicates pretransitional ferroelectric soft-mode behavior. A notable feature of these data is the overlap of the high frequency falloff of the $\varepsilon_A''(\omega)$ curves, also a tendency in *Fig. 4* for the N phase (80/20 and 70/30 mixtures). In a PCG-like model, this overlap would indicate that the effective $R_{NF}$ associated with the soft mode is independent of $T$. Then, since $\tau_o = R_{NF}C_I$, and $\varepsilon_A''(f_o) = \varepsilon_A'(f)/2 \propto C_I$, we would have $\varepsilon_A''(f_o) \propto \tau_o = 1/f_o$, which is the case inside the orange circle.

(*iv*) Additional information about the nature of ionic charge transport in the PCG model comes from SmAP$_F$ experiments where ion current bumps, due to ions crossing the $N_F$ layer, appear only for $V > V_{sat}$ [4], the regime where $E_{NF} = (V - V_{sat})/d$ shown in *Fig. 2B*. With $|V| < V_{sat}$, the field in the $N_F$ layer is small and the ions are not mobile, contributing little to the current through the $N_F$

(*v*) Ionic species in the $N_F$ may partially screen the polarization charge, an effect tending to increase $\xi_P$ that has been considered under a variety of conditions [40,50,54]. However, an effect that has not been considered in the literature, which is particularly relevant to ferroelectric nematics due to their high polarization, is that the polarization charge can screen the ions. This is, in essence, what is happening in (*i*) and (*iv*) above. We consider the cell of *Fig. 1* with an equimolar concentration of + and – ions in the $N_F$. Suppose that a voltage is applied but initially $\psi$ is clamped at $\psi = 0$. The field in the $N_F$ will separate the ionic charge into double-layers at the surfaces, each of thickness $\xi_D$, the Debye screening length. If $d \gg 2\xi_D$, the applied field will be screened to near zero in the cell center but will be non-zero in the double-layers. In an $N_F$ with $P$ of the order of $\mu C/cm^2$, the condition $\xi_D \gg \xi_P$ will always be met, in which case, if $\psi$ is released, reorientations of $P$ induced by the field in the double layer will reduce this field, causing the double layer to expand into the cell center. The ongoing process will produce a cell like that of *Fig. 1* with zero field in the $N_F$, but with the ions randomly distributed except possibly at PSKs on the surface of thickness $\xi_P$. The ions will diffuse to an on-average uniform, zero net charge distribution. This scenario is consistent with the experimental observations related in in (*i*) and (*iv*) above.

*The need for publishing complete data sets*
Based on the preceding discussion, we would maintain that experimental data sets comprising measurements of $\varepsilon_A(\omega)$ of the $N_F$ phase are fundamentally incomplete unless they include key parameters such as the operative values of $A$ and $d$ for parallel-plate cells and details of the electrode geometry necessary for deriving the raw impedance data $Z_\omega$ from $\varepsilon_A(\omega)$. It is also essential to report information on the nature and preparation of the interfacial layers at the $N_F$ / electrode interfaces that is as complete as possible. Indeed, it would be very useful if such addenda were provided for currently available data sets, *e.g.*, those in Refs. [9, 13-24]. In the PCG frequency regime, Re$Z_\omega$ will determine $C_I$. Im$Z_\omega$ will peak at a single-relaxation frequency, $\omega_R = 1/\tau_o$. With $C_I$ and the measured $\tau_o$, we have $R_{NF} = \tau_o/C_I$, and from $R_{NF}$ we get $\rho_{NF} = AR_{NF}/d = \gamma/P^2$, enabling measurement of $\gamma$ for known $P$. Without access to all of the relevant experimental parameters, it is not possible to verify or further analyze with any confidence the results of dielectric measurements of high polarization ferroelectric liquid crystals.

## *Discussion*

In modeling the dielectric response of liquid crystals, it is typically assumed that the impedance of the LC is large compared with that of its interfacial layers. In this limit, the capacitance of the insulating layers $C_I \to \infty$, meaning the voltage applied to the cell all appears across the LC. A typical ferroelectric nematic in a typical cell presents the qualitatively opposite limit: nearly all of the voltage applied to the cell appears across the interfacial layers. A particular example of this situation, where the interfacial layers are insulating, thin, and therefore highly capacitive, has been analyzed here. Two basic effects, which originate from the screening of electric fields by polarization charge and cannot be ignored in liquid crystals with high polarization, control the emergent physical picture of $N_F$ electrodynamics: (*i*) the elimination of the electric field from the bulk LC due to polarization charge and the resulting confinement of polarization charge to the interfaces, and (*ii*) the drastic reduction of the applied electric field within the ferroelectric volume. For capacitive interfacial layers, these effects couple polarization reorientation (a Goldstone process) to charging of the interfacial capacitance, making these layers the principal energy storage mechanism of the Goldstone mode [4] and leading to the polarization-capacitance Goldstone (PCG) model.

A central prediction of the PCG model is that the measured capacitance of cells in the $N_F$ phase will be $C_I$, the capacitance of the insulating layers, which provide the dominant impedance at low frequency. The experiments by Nishikawa *et al.* [19,20] are in accord with this prediction, showing that with a chosen cell type having reproducibly made insulating layers, the dominant cell impedance is capacitive and that the capacitance measured at low frequency does not depend on the details of the $N_F$ liquid crystal. The cell behaves simply like a resistor, with the capacitance effectively determined by the permittivity and thickness of the insulating layers. This insensitivity to conditions in the $N_F$ is somewhat surprising but may open the door for using the $N_F$ phase to probe the *in-situ* structure and electrical characteristics of the interfacial layers.

After this extended discussion of sandwich cells, it is worth pointing out that a similar analysis could be carried out on cells with electrode structures configured to produce in-plane electric fields. For example, cells with metal electrodes evaporated on only one of the glass plates and $N_F$ between and over the electrodes are frequently used to experiment with $N_F$ materials. In sandwich cells, the electrode plates are assumed to be planar, parallel, isopotential surfaces and for the PCG model based on a 1D variation of the polarization orientation $\psi(x)$, the spatially intermediate isopotentials will always be similarly planar and parallel. This leaves no path for electric field lines or current flow to get around high impedance layers, thus giving $R \propto d/A$, and $C \propto A/d$, for example. In the $N_F$ phase, polarization screening affects every path the same way. In the in-plane case, though, some field lines paths pass through the $N_F$ material and carry the current while others pass through glass or air, and some through all three, the overall combination of which will determine the shape of the isopotentials. This produces differences between the electrical effects in sandwich and in-plane cells that are unique to ferroelectric LCs.

The results presented here for purely capacitive interfacial layers can be readily extended to more complex interfacial electric behavior. Returning to Eqs. 11,12, this is achieved simply by replacing the imaginary capacitive impedance $1/i\omega C = Z_C$ with a complex impedance $Z_I$, which represents more general interface electrical behavior. For example, a simple leaky interface could be modeled as a parallel interfacial resistor/capacitor combination by $Z_I = R_I Z_C/(R_I + Z_C)$, etc.

In summary, we have shown that the appearance of fluid ferroelectricity produces a dramatic, qualitative transformation of the fundamental electrical structure of a planar aligned LC cell. This transformation can be understood by comparing the electric structure of an N phase to that of the $N_F$. Cells of both of these phases have bulk insulating LC bounded by non-ferroelectric, largely insulating, thin layers at the planar electrode surfaces. These interfacial capacitances are in series with the bulk LC, a much lower capacitance, in which combination the applied voltage appears almost entirely across the lower capacitance LC. Measurement then yields the capacitance of the LC layer with small corrections due to the interfaces. In the $N_F$ phase, a static field induces torque on the polarization. The resulting reorientation of $P$ is in the direction that reduces the field in the LC, resulting in the field in the LC going to zero, and remaining small even in the dynamic case (ferroelectric superscreening). Now the voltage applied to the cell is almost entirely across the interfacial capacitors. The LC has gone from being the high impedance element in the N cell to being the low impedance element in the $N_F$ cell. Understanding and applications of the electrical behavior of the $N_F$ must be sought on this basis.

## *APPENDICES*

*Appendix A – From the Goldstone mode to the PC Goldstone mode* – The description of the Goldstone mode applicable to low-polarization FLC phases and its dielectric constant prediction can be obtained from the torque density balance equation, following Carlsson *et al.* [33]

$$\gamma(x,t) = PE_{NF}(t)\cos\psi(x,t) - K\psi_{xx}(x,t) = P[E_a(t) + E_P(x,t)]\cos\psi(x,t) - K\psi_{xx}(x,t), \qquad (21a)$$

with

$$E_P(x,t) = P/\varepsilon_{LC}[\sin\psi(-d/2,t) + \sin\psi(x,t)]. \qquad (21b)$$

Here $E_a(t) = V(t)/d$, is the externally applied field, which is also the field in the N$_F$. Referring to *Fig. 1*, this implies the limit $C_I \to \infty$ has been taken, meaning that the polarization charge at the electrode interfaces is completely screened by free charge. This eliminates the electrostatic restoring force of Eq. 17b. $E_P(x,t)$ is the field in the N$_F$ due to polarization space charge [26], which can be ignored for sufficiently small $P$ [33]. With these assumptions, $E_{NF}(t) = V(t)/d$ and for $\psi(x,t) \ll 1$ we have

$$\gamma\dot\psi_q(t) = PE_{NF}(t) - Kq^2\psi_q(t) \qquad (21c)$$

with the solution for $\psi_q(t) \propto \exp(i\omega t)$

$$\psi_q = (PE_{NF}/Kq^2)(1 + i\omega\gamma/Kq^2)^{-1}. \qquad (21d)$$

In a cell, the polarization field deformation $\psi(x) = \psi_q \cos qx$, with $q = \pi/2d$, is the splay-deformation Fourier mode of the director/polarization field that has the smallest $q$ that satisfies the boundary conditions $\psi(0) = \psi(d) = 0$. The spatial average of $P_x$ over this deformation gives the dielectric constant (see Eqs. 37-39 in Ref. [33])

$$\varepsilon_A(\omega) = \langle P_x\rangle/E_{NF} = P\langle\psi\rangle/E_{NF} = \varepsilon_{LC}[(P^2/\varepsilon_{LC})/Kq^2](1 + i\omega\gamma/Kq^2)^{-1} = \varepsilon_{LC}[1/\xi_P^2 q^2](1 + i\omega\gamma/Kq^2)^{-1}. \quad (21e)$$

Here, as in the discussion following Eq. 16, $\varepsilon_{LC}$ is the bare, polarization-free, N$_F$ dielectric constant, and we have substituted in the polarization penetration length $\xi_P$ from Eq. 1. The term in square brackets is the ratio of the polarization self-energy density, $P^2/\varepsilon_{LC}$, to the Frank elastic energy density, $Kq^2$, which ratio, from Eq. 1 is a very large number (~$10^4$) in the case of the N$_F$ for sandwich cell electrodes separated by a $d$ of several microns. This result would appear to predict the potential for the N$_F$ to exhibit a large, Goldstone-mediated $\varepsilon_A(\omega \to 0)$. However, Eq. 21c-e as written are not applicable to the N$_F$ because including the polarization term increases the energy cost of elastic deformation, as discussed in numerous publications [1,3,50-55].
50,51,52,53,54,55]
Specifically, in the high-polarization limit relevant to the N$_F$, if $P^2/\varepsilon_{LC} \gg Kq^2$, then it follows from Eqs. 21a,b that, $P^2/\varepsilon_{LC} \sim PE_P(x,t) \gg Kq^2$ and the $E_P(x,t)$ term in Eq. 19b is dominant and cannot be ignored. In this regime, Eq. 21 gives solutions that are solitonic rather than sinusoidal [26], as shown in *Fig. A1*, with $\psi(x)$ becoming spatially uniform away from surfaces, approximating the $q = 0$ mode, with the cell free of space charge and electric field. Any spatial variation of $\psi$ is confined to polarization-stabilized kinks (PSKs) [26], lines in 2D or sheets in 3D of thickness $\xi_P$, located at places where boundary conditions or topology require deviations from uniformity.

Application to the present case of a planar-aligned $N_F$ cell, in which $\xi_P \ll d$, yields a polarization field which is spatially uniform in the bulk with an orientation described by the single scalar variable $\psi(t)$, and that has surface layers of thickness $\xi_P$ that transition the field to the preferred surface orientations. Example responses of the polarization to an electric field applied along $x$ are sketched in *Fig. A1C*.

However, having PSKs with significant reorientation at the surfaces is an unlikely situation in an $N_F$ since, as calculated in Eq. 17, the torques required in the $N_F$ phase for the surfaces to maintain $d\psi(x)/dx \sim 1/\xi_P$ are orders of magnitude larger than typical LC surface interactions, in which case the PSKs will pop out and a uniform $\psi(t)$ will fill the LC except for some surface layer of nanometer thickness. This is the liquid crystal structure depicted in *Fig. 1*. The LC/electrode interfaces will exhibit some remnant $\psi$-dependent surface interaction, which will depend on the details of the interface structure.

*Appendix B – Dielectric constant of an isolated $N_F$ slab* – Consider a ferroelectric slab having a *P*-field of freely rotatable ferroelectric polarization vectors in a medium of bare dielectric constant $\varepsilon$. Present is a small electric field *E* normal to the slab, inducing $\psi(E)$, a small reorientation of *P*. For $E = 0$, *P* is stabilized at $\psi = 0$ by the electrostatic energy $U = \frac{1}{2}(P^2/\varepsilon)\psi^2$ and resulting restoring torque $\mathcal{T}_R = -(P^2/\varepsilon)\psi$, produced by the surface polarization charges $\sigma = \pm P\psi$. Addition of the applied field-induced torque, $\mathcal{T}_A = PE$, results in a net torque $\mathcal{T}_N = \mathcal{T}_A - \mathcal{T}_R = PE - (P^2/\varepsilon)\psi = PE_s$ where $E_s$ is the electric field in the slab. This gives for a polarization charge $Q_\omega = PA\psi_\omega$ on the surface:

$$\gamma d\psi(t)/dt = PE_s = PE - (P^2/\varepsilon)\psi \qquad (22a)$$

or

$$(\gamma/P^2)i\omega\, Q_\omega/A = E_{s\omega} = E_\omega - Q_\omega/A\varepsilon \qquad (22b)$$

$$Q_\omega/A = \varepsilon E_\omega[1 + i\omega(\gamma\varepsilon/P^2)]^{-1} = \varepsilon E_\omega[1 + i\omega\rho_{NF}\varepsilon]^{-1} \qquad (22c)$$

The dielectric constant of the slab due to *P* is

$$\varepsilon_{Ps} = (Q_\omega/A)/E_{s\omega} = \varepsilon E_\omega[1 + i\omega\rho_{NF}\varepsilon]^{-1}/\{E_\omega - Q_\omega/A\varepsilon\} \qquad (22d)$$

$$\varepsilon_{Ps} = (Q_\omega/A)/E_{s\omega} = \varepsilon[1 + i\omega\rho_{NF}\varepsilon]^{-1}/\{1 - [1 - i\omega\rho_{NF}\varepsilon]^{-1}\} = \varepsilon/i\omega\rho_{NF}\varepsilon = \varepsilon/i\omega R_{NF}C_s \,. \quad (22e)$$

Adding the capacitive current through $C_s$ adds the term $C_s E_{\omega s} d_s/A = \varepsilon_s E_{\omega s}$ to *D*, giving $\varepsilon_s = \varepsilon + \varepsilon_{Ps}(\omega)$, the same result obtained in Eqs. 16 a,b. As the frequency is lowered, the applied electric field is increasingly screened away inside the slab. The key condition for this to take place is that the electrical torque $\mathcal{T}_N$ be the only torque on *P*. If there is an additional restoring torque on *P*, then $\varepsilon_{Ps}$ will saturate at a finite real value.

The role of the interfacial capacitors $C_I$ can be understood by envisioning that $E_\omega$ in Eq. 22 is produced by a pair of positive and negatively charged planar electrodes on either side of the slab. The external field $E_\omega = Q_\omega/\varepsilon A =$ is required to induce reorientations $\psi_\omega = Q_\omega/P$ and doing so costs energy $U = \frac{1}{2} \varepsilon |E_\omega|^2 v$, where $v$ is the volume between the electrodes. In the absence of other torques on $P$, this is the only reversible energy cost of changing $\psi$. Reducing the gap between the electrodes and the slab (and therefore increasing $C_I$) reduces $v$ and therefore reduces $U$, making reorientation easier, a process that can be continued until $d_I$ reaches molecular scale, in which case the apparent $\varepsilon_A(\omega \rightarrow 0)$ will be large.

*Appendix C – Recent $N_F$ examples* – Broadband measurements of the dielectric constants of the $N_F$ phase of RM734 [56] and DIO [57] have recently been published which provide the first full dielectric spectra of ferroelectric nematics as a function of frequency, sample thickness and temperature. These two data sets provide essentially identical, direct, model-independent evidence that the claimed *giant bulk-phase capacitance* of the $N_F$ phase [9, 13–24] does not exist in either RM734 or DIO. The results presented above have shown that the capacitive characteristics of $N_F$ cells observed at low frequency, *i.e.*, the apparent dependence of the measured capacitive $\varepsilon_A(\omega)$ on cell thickness at low frequency, can be understood only by introducing surface capacitive layers. Thus, references [56,57], which present impressive sets of measurements of the dependence of $\varepsilon_A(\omega)$ in RM734 and DIO on sample thickness for both the low ($\omega\tau_o < 1$) and high ($\omega\tau_o > 1$) regimes, are of particular relevance to this discussion. Data from Ref. [56] (Figs. 5 and S2, reproduced here as *Fig. C1a-d*), are important because they contribute key measurements of the thickness dependence of $\varepsilon_A(\omega)$ and $Z(\omega)$ in the $\omega > 2\pi f_r$ regime to the data pool, enabling comparison of the low and high frequency regimes (note that $\tau_o = 1/2\pi f_r$ in the notation of Ref. [56]). Ref. [56] extracts the apparent dielectric constant $\varepsilon_A(\omega)$ from the measured impedance $Z(\omega)$ under the homogeneous dielectric assumption shown in Eq. 13a,

$$\varepsilon_A(\omega) = (d/A)[1/i\omega Z(\omega)].$$

As pointed out above, this measured $\varepsilon$, in the geometry of *Fig. 1* and with no DC field applied, is an $\varepsilon_{xx}$. At low frequency, the measured $Z(\omega)$ values in Ref. [56] are substantially independent of sample thickness, $d$ [*Fig. C1c,d* (*blue circles*), Ref. [56] Fig. S2], evidence of a dominant contribution by the interfacial impedance for $\omega < \omega_r$, and yielding, according to Eq. 13a, an apparent $\varepsilon_A(\omega)$ that is therefore proportional to $d$ [*Fig. C1a,b* (*gray circles*), Ref. [56], Figs. 5,8]. This is as expected in the PCG model (Eq. 15a), although in the case of Ref. [56] the interfacial impedance is not simply capacitive but rather some RC combination, likely due to resistive leakage through the nanoscale-thick capacitive layer at the $N_F$/bare gold interface [58].

For $\omega > \omega_r$, in contrast, $Z(\omega)$ is proportional to $d$ [*Fig. C1c,d* (*red circles*), Ref. [56] Fig. 5], producing an $\varepsilon_A(\omega)$ that is independent of $d$ [*Fig. C1b* (*green circle*), Ref. [56] Fig. 5]. This behavior indicates a second key result, namely that in the $\omega\tau_o > 1$ regime, the measured $\varepsilon_A(\omega)$ is behaving like a bulk dielectric property, specifically that of a *bulk resistance*: $\varepsilon_A(\omega) = \sigma_{NF}/i\omega$, [**Fig. C1b** green line] wherein: (*i*) Im$\varepsilon_A(\omega)$ varies as $1/\omega$, and, importantly, (*ii*) $\sigma_{NF}$ is the effective, thickness-independent bulk conductivity of the N$_F$ layer.

Referring back to the PCG model, if in Eqs. 3 to 8 we let $C_I \to \infty$, then $Z_I \to 0$, $V_{sat} \to 0$, we are guaranteed to be probing the bulk impedance, and to be in the $\omega\tau_o > 1$ regime. From Eq. 4 we have in this limit

$$I(t) = AP[d\psi(t)/dt] = AP[PE_{NF}/\gamma] = AP[(P/\gamma d)V_{NF}] = (A/d)(P^2/\gamma)V(t) = (A/d)(\sigma_{NF})V(t) = V(t)/R_{NF},$$

with $\sigma_{NF} = P^2/\gamma$. Thus Refs. [56,57] present additional experimental confirmation of the predicted PCG behavior in RM734, showing that Eq. 4 describes the behavior over a four decade frequency range ( $2.5 < \log f < 6.5$). Once *f* is large enough that the interface impedance drops out, RM734 in the N$_F$ phase behaves like a bulk resistive/semiconducting slab of conductivity $\sigma_{NF}$ in this frequency range. Using **Fig. C1b** to obtain a measured conductivity gives $\sigma_{NF} \sim 10^{-3}$S, which, for a $d = 7$ μm thick cell of area $A = 5$ mm x 5 mm cell as used in Refs. [56,57], gives a cell resistance value $R_{NF} \sim 300$ Ohms, and accurately proportional to *d*, given the constancy of $\varepsilon_A(\omega)$.

A common feature of single-relaxation systems such as in Langevin-Debye dielectric relaxation or in the Goldstone mode considered here is domination of reactive (real ε) behavior at low frequencies and of dissipative behavior (imaginary ε) at high frequencies. The relaxation frequency $\omega_r$ is a crossover condition where these balance in magnitude, the relaxation time τ being a ratio of reactive over dissipative coefficients. This is exemplified in the PCG mode wherein the relaxation frequency $\omega_r = 1/\tau_o = 1/R_{NF}C_I$ from Eq. 8 is the crossover condition $R_{NF} = 1/\omega_r C_I$, where the magnitudes of the high frequency bulk resistive and low frequency interface capacitive impedances are the same. Since $R_{NF}$ is proportional to *d*, and $C_I$ is independent of *d*, this condition requires that $\omega_r$ vary as $1/d$ as in Fig. 8 of Ref. [56], and in Fig. 8 of Ref. [57]), a dependence that is not a result of any molecular processes but simply emerges from the geometry of a plot having $\varepsilon_A(\omega)$ vary as *d* on the reactive side and be independent of *d* on the dissipative side. This is graphically clearest in the N$_F$ phase data in Fig. 6 of Ref. [57], but the fits to $\omega_r \sim 1/d$ behavior in Fig. 8 of Ref. [56], and Fig. 8 of Ref. [57] show that is a common feature of all the data in these papers. These fits therefore show quantitatively that the only substantial contributions to $\varepsilon_A(\omega)$ for $\omega < \omega_r$ have $\varepsilon_A(\omega) \propto d$, and that the only substantial contributions to $\varepsilon_A(\omega)$ for $\omega > \omega_r$ have $\varepsilon_A(\omega)$ independent of *d*, an inference that can be made without reference to any specific model. These fits eliminate the possibility of there being any substantial bulk capacitance in these materials, again independent of any specific model.

*A*CKNOWLEDGEMENTS

This work was supported by NSF Condensed Matter Physics Grants DMR 1710711 and DMR 2005170, and by Materials Research Science and Engineering Center (MRSEC) Grant DMR 1420736.

*Figures*

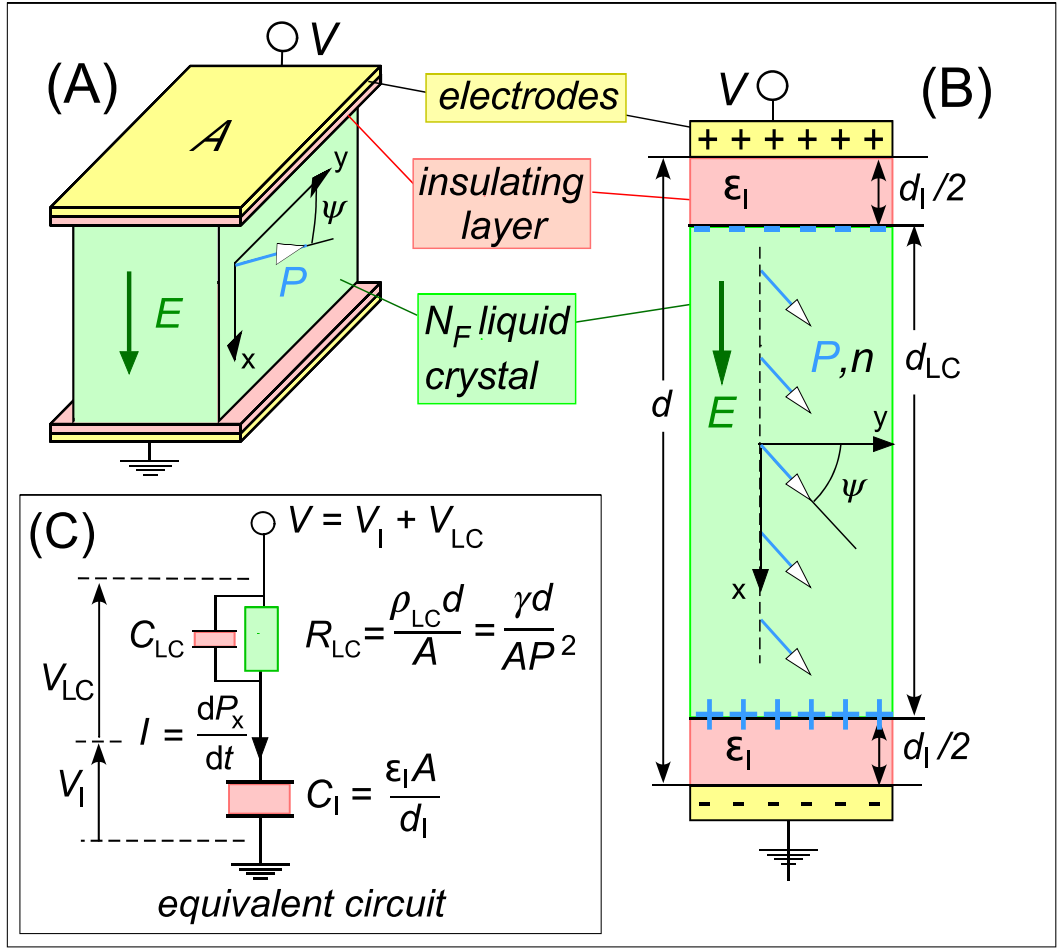

*Figure 1*: (*A*) Geometry of a planar-aligned $N_F$ cell, of area *A*, used for dielectric constant measurement and exhibiting a PCG mode. (*B*) Cross-section of the cell, showing the LC and insulating layers, $N_F$ polarization, polarization orientation $\psi(t)$, free charge (black), and polarization charge (blue). Electrostatic self-screening causes the polarization field to be uniform, with polarization charge expelled to the LC surfaces. This diagram is for the case of large *P*, where electrostatic stabilization of uniform *P* has overcome any Rapini – Papoular surface energy, eliminating the polarization-stabilized kinks of Appendix *Fig. A1* at the $N_F$ surfaces and depositing the polarization charge within a molecular-scale length of the $N_F$ surfaces. (*C*) Electrical equivalent circuit of the cell. The $N_F$ layer behaves electrically like a resistor with a resistivity $\rho_{LC} = \gamma/P^2$. This layer is in series with the insulating, interfacial layers of capacitance $C_I$. The $I_\omega$-$V_\omega$ characteristic of the circuit impedance, $Z_\omega = R_{LC} + 1/i\omega C_I$, gives the frequency dependence of the complex measured permittivity $\varepsilon_A(\omega)$. $C_{LC}$ represents the "bare" capacitance of the $N_F$, coming from its dielectric response in absence the effects due to *P*. If $d_{LC} \gg d_I$, then $C_{LC} \ll C_I$, and in the $N_F$ phase we have $Z_{C_{LC}} \gg Z_{C_I}, Z_{R_{LC}}$, in which case we can ignore the effects of $C_{LC}$ in calculating the dielectric response. Adapted from Ref. [4].

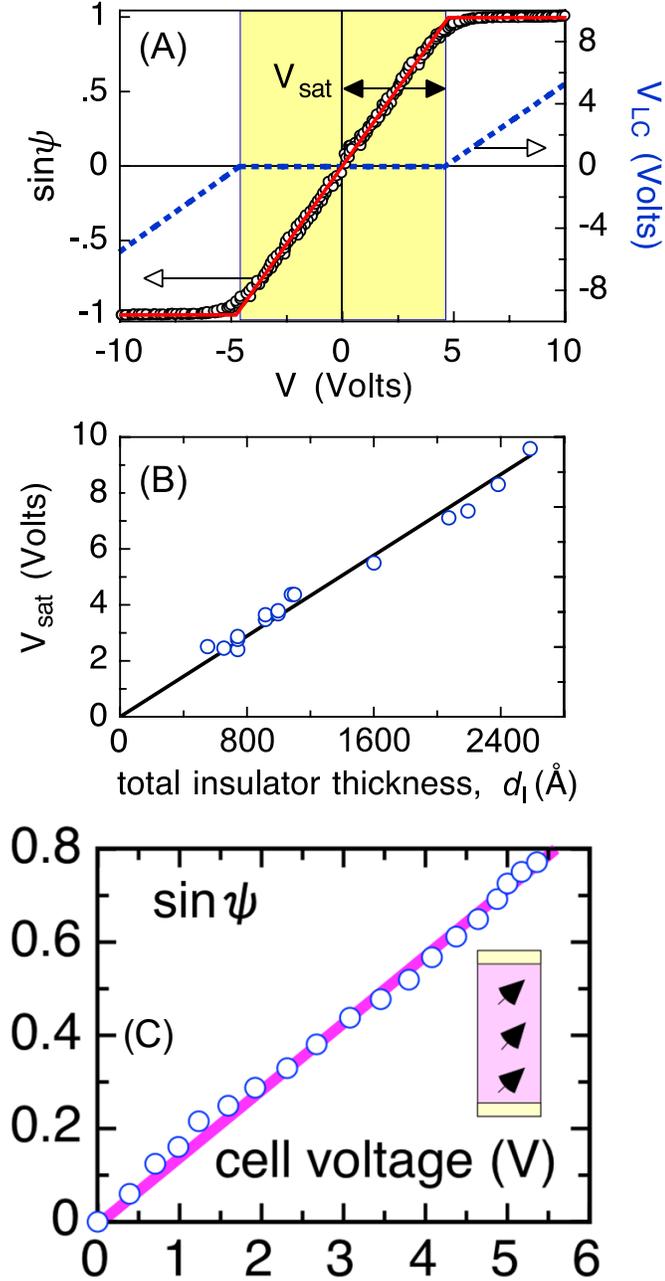

*Figure 2*: (*A*) Electric field-induced, optically measured, change in polarization orientation of a SmC* FLC with $P \approx 0.23$ µC/cm$^2$ (black symbols), reproduced from Ref. [22] with permission, plotted along with the PCG model result of Eq. 2a (red), reproduced from Ref. [59] with permission. Here $\psi$ is the orientation of *P* as defined in *Fig. 1*. Also shown is the expected voltage, $V_{LC}$, appearing across the LC under quasi-static conditions. (*B*) Measured saturation voltage, $V_{sat}$, vs. the total thickness of the two evaporated SiO$_2$ insulating layers, $d_I$, reproduced from Ref. [8] with permission. The slope is $P/\varepsilon_I$. (*C*) Electric field-induced, optically measured, change in polarization orientation about the layer normal of a SmAP$_F$ FLC with $P \approx 0.4$ µC/cm$^2$, reproduced from Ref. [22] with permission. Here $\psi$ is the orientation of *P* as defined in *Fig. 1*.

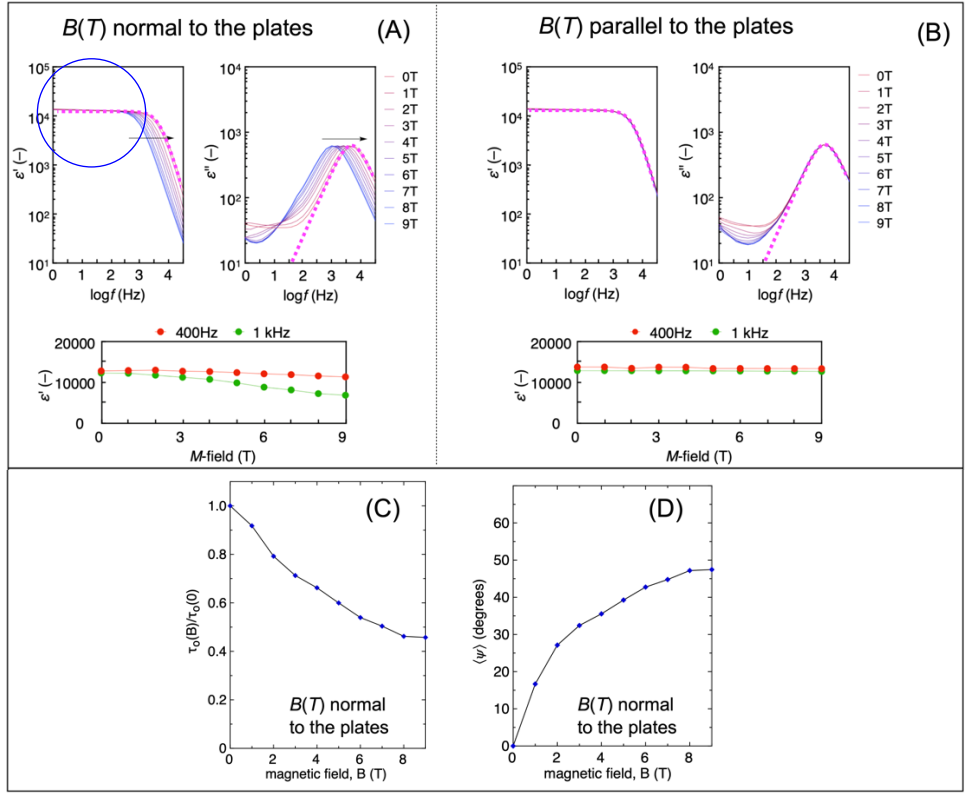

*Figure 3*: Dielectric spectra, reproduced from Ref. [**19**] with permission, of a planar-aligned cell of an $N_F$ phase in an applied magnetic field. (*A*) Magnetic field normal to the plates. The scaled PCG model for Re $Q_\omega$ (Eq. 10), plotted in magenta for *T* = 44°C, describes the major features of the frequency dependence of PCG feature in any these $N_F$ spectra quite well. (*B*) Magnetic field parallel to the plates. In this case, the spectra change little despite the very large magnetic field because *P* is already parallel to the plates at *B* = 0. (*C*) With *B* normal to the plates, the relaxation time, $\tau_o = R_{LC}C_I$, changes, which, according to the PCG model, can be used with Eq. 8 to obtain $\langle\psi\rangle$, the field-induced tilt angle of *P*, with the result presented in (*D*). At *B* = 9 T, which is ~ 60–90 times the typical magnetic Freedericksz threshold for such a reorientation in a non-ferroelectric nematic, the $N_F$ orientation is still only $\langle\psi\rangle$ ~ 45°. This result confirms that $\langle\psi\rangle$ is strongly stabilized in the planar orientation by the cell electrostatics, as predicted by the PCG model. A key feature of this data is that $\varepsilon_A'(0)$, the low-frequency asymptotic value of $\varepsilon_A'(\omega\to 0)$ [and therefore the peak value of $\varepsilon_A''(\omega = 1/\tau_o) = \varepsilon_A'(\omega)/2$ in Eq. 15] exhibits little dependence on magnetic field strength *B* (blue circle) and therefore little dependence on the $N_F$ orientation $\langle\psi\rangle$. This can be readily explained by, and is direct evidence for, the PCG model: in an RC circuit, for $\omega\tau_o \ll 1$, the cell impedance is dominated by $C_I$, which is apparently independent of $\langle\psi\rangle$. (*NOTE*: Comparison of this Figure to others in [19] and to *Fig. 4* shows that the $\varepsilon''$ scale in *A* and *B* is actually 10 times larger than indicated in this figure.) *A*,*B* adapted from Ref. [19] (blue ellipse added). *C*,*D* extracted from *A*.

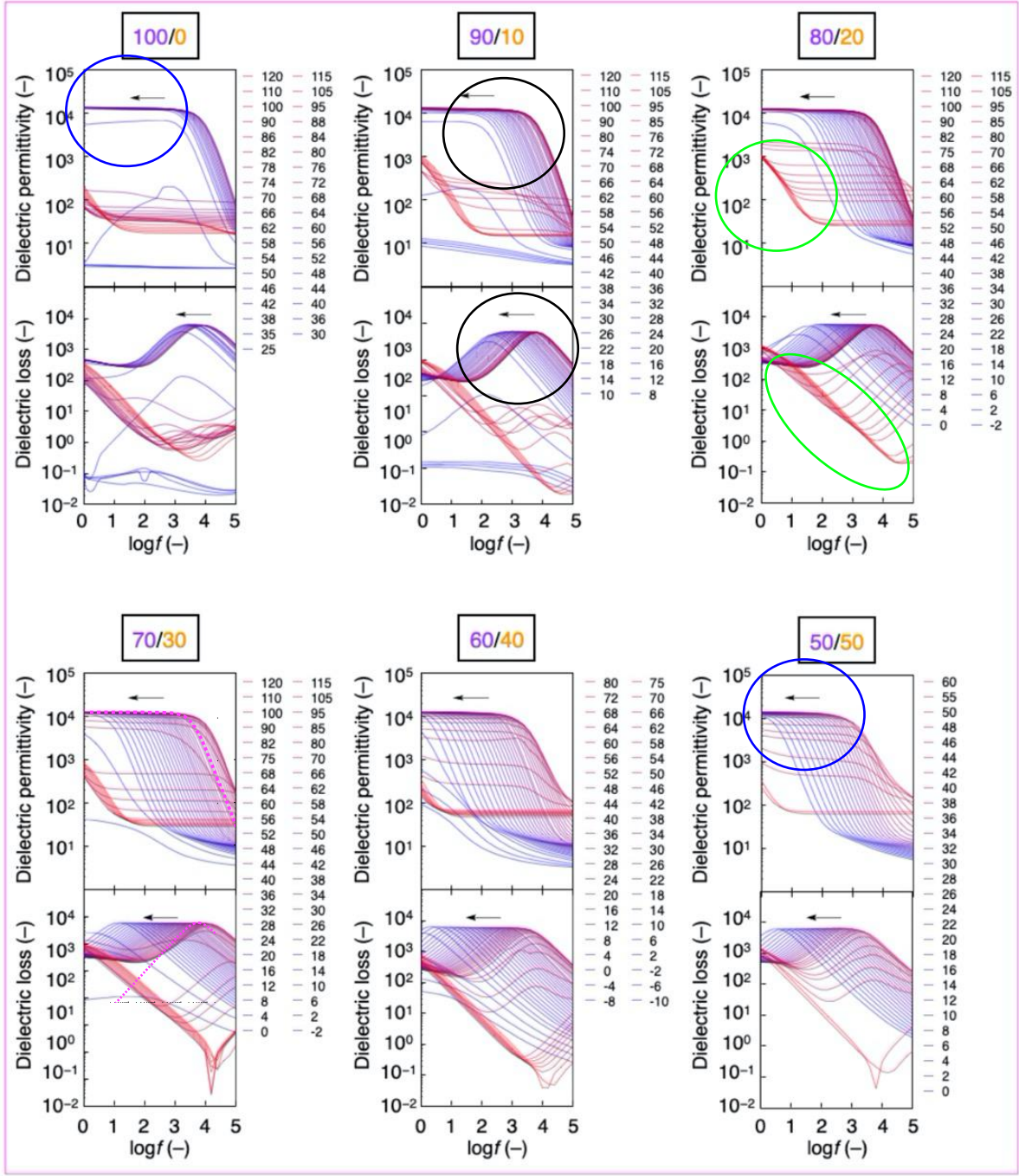

*Figure 4*: Scans at different temperatures, reproduced from Ref [20] with permission, of the dielectric spectra of a series of binary mixtures of DIO homologs having the N$_F$ phase. The PCG mode features (*e.g.*, black circles) are similar to those of *Fig. 3*. The scaled PCG model for Re $Q_\omega$ (Eq. 10), plotted in magenta for one temperature in the 70/30 case, describes the major features of the frequency dependence of all of these N$_F$ spectra quite well. A notable feature of these data is that, while the $\tau_o = R_{LC}C_I$ values vary over orders of magnitude with $T$, $\varepsilon_A'(\omega \to 0)$ and the peak value of $\varepsilon_A''(\omega = 1/\tau_o) = \varepsilon_A'(0)/2$, are, remarkably, nearly the same for all of the different

temperatures, $\tau_o$ values, and material compositions. According to the PCG model (Eq. 15), the predicted low-frequency asymptotic value of is $\varepsilon_A'(\omega \to 0) = dC_I/A$. The Experimental Section of Ref. [20] suggests that these measurements were made in cells with the same values of $d$ and $C_I$. The invariance of the low-frequency response with mixture composition (blue circles) would appear to confirm this to be the case. The variation of $\tau_o$ indicates that $\rho_{LC} = AR_{LC}/d = \gamma/P^2$ changes strongly with temperature in these mixtures. The green ellipses highlight the low-frequency ionic current feature, which disappears in the $N_F$ phase because of polarization charge screening of the $E$ field in the $N_F$. Adapted from Ref [20] (colored ellipses added).

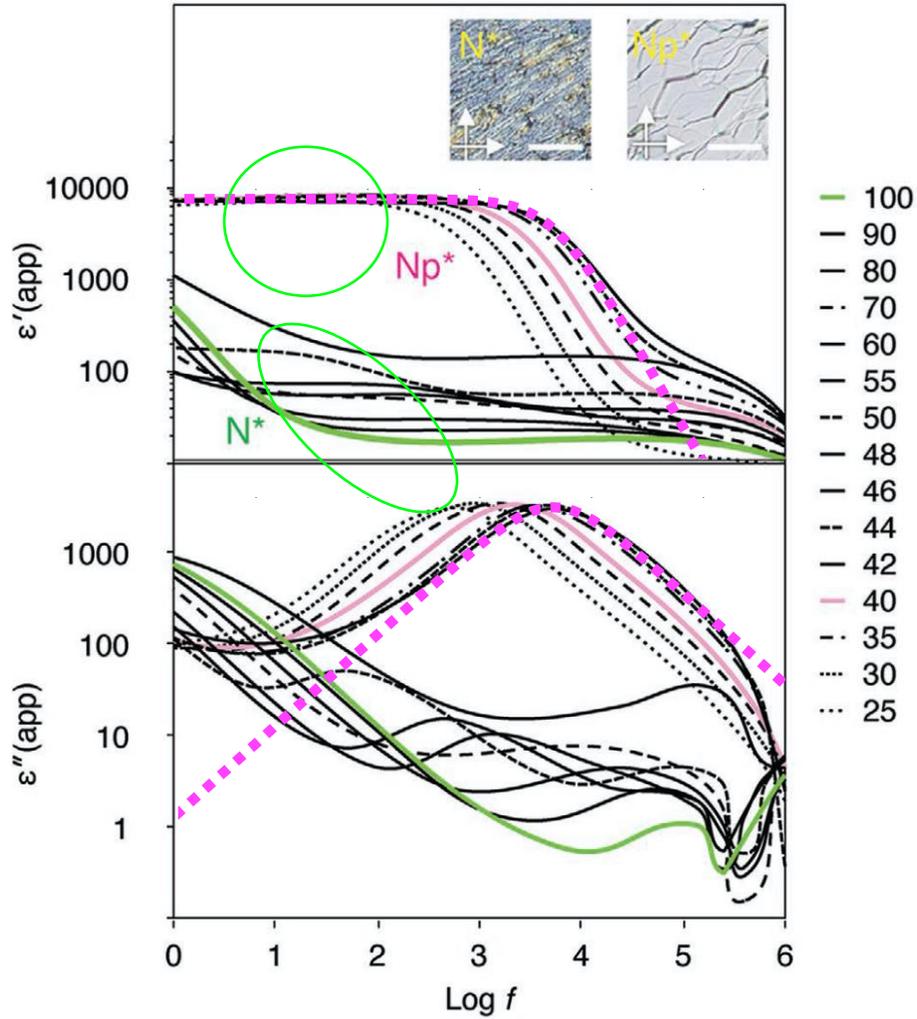

*Figure 5*: Dielectric spectra, reproduced from Ref. [15] with permission, showing $\varepsilon_A(\omega)$ values obtained on cooling DIO with a chiral dopant, making a helical polar nematic of the $N_F$ phase, here termed $N_P^*$. The chiral $N_F$ cells exhibited Grandjean textures as shown in the image, which for dielectric purposes means that these cells are planar, with *P* parallel to the plates, the preferred $N_F$ orientation. The scaled PCG model for Re $Q_\omega$ (Eq. 10), plotted in magenta for *T* = 44°C, describes the major features of the frequency dependence of the permittivity in the $N_P^*$ phase quite well. The green ellipses highlight the low-frequency ionic current feature, which disappears in the $N_F$ phase because of polarization charge screening of the *E* field. Adapted from Ref. [15]. Copyright Wiley-VCH GmbH. (Magenta curves and green ellipses added.)

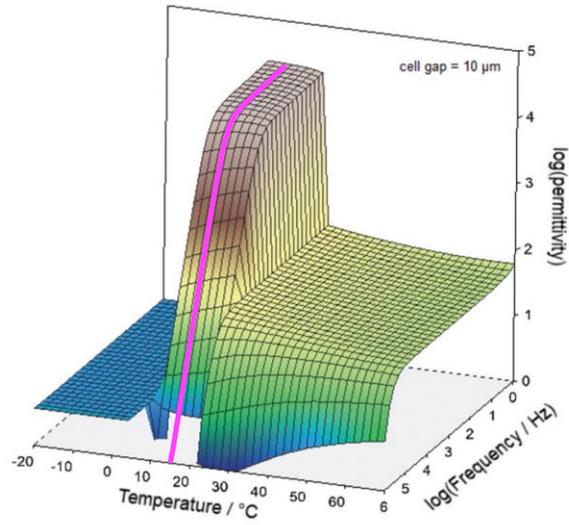

*Figure 6*: Dielectric spectra, reproduced from Ref [14] with permission, showing $\varepsilon_A'(\omega)$ values obtained on cooling of a single-component mesogen that is a ferroelectric nematic in the temperature range 20°C ≳ $T$ ≳ 10°C. Sample orientation was not experimentally determined but is likely planar. The scaled PCG model for Re $Q_\omega$ (Eq. 10), plotted in magenta for $T$ = 15°C, describes the major features of the frequency dependence of the permittivity in the N$_F$ quite well. Adapted from Ref [14]. Copyright Taylor & Francis. (Magenta curve added)

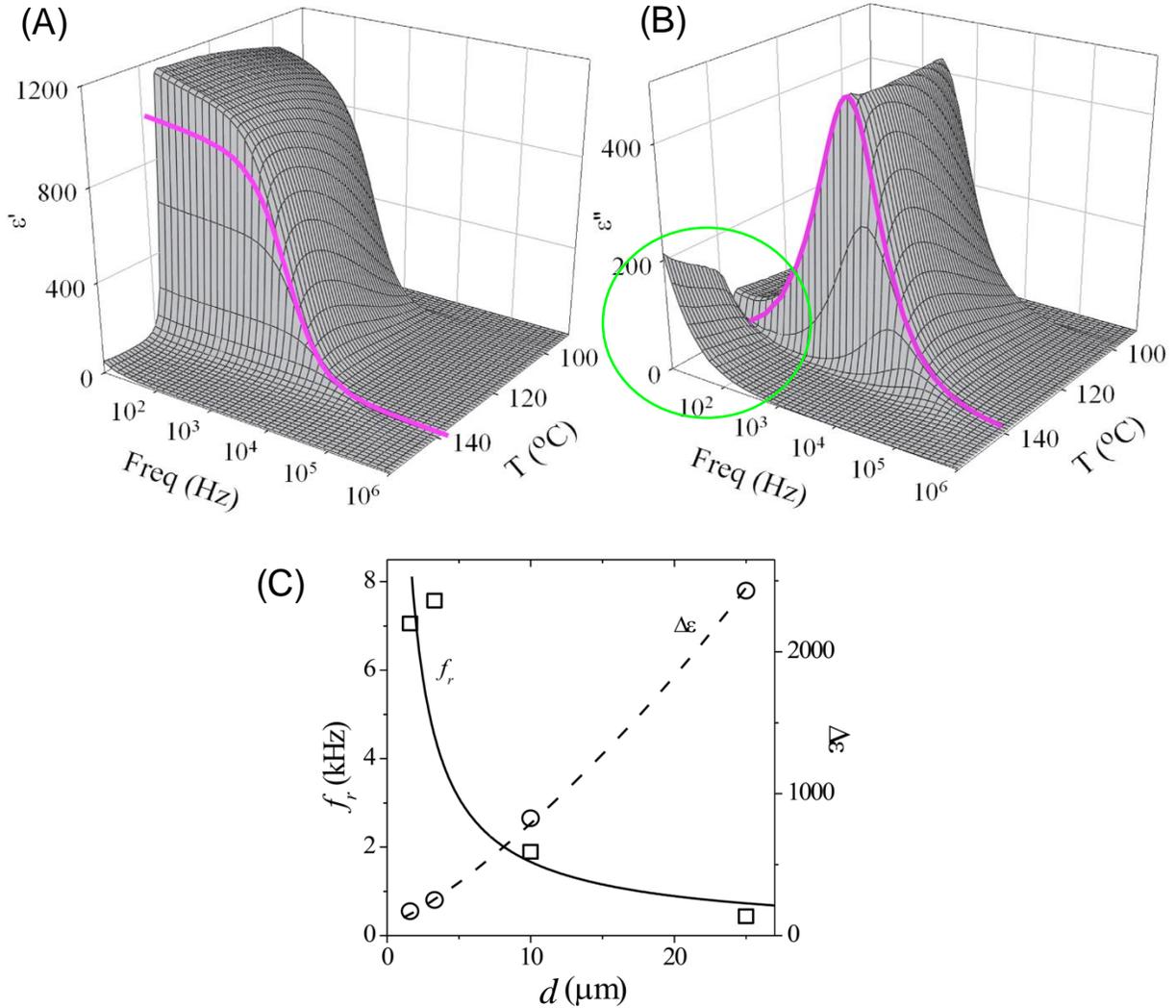

*Figure 7*: Dielectric spectra, reproduced from Ref [23] with permission, obtained on cooling the bent-core mesogen W586 into the SmAP$_F$ phase. The polarization varies from $P = 0.1$ μC/cm² at $T = 138$°C to $P = 0.35$ μC/cm² at $T = 100$°C in the SmAP$_F$ phase. The sample has bookshelf layering with $P$ parallel to the cell plates. (A,B) The scaled PCG model for Re $Q_\omega$ (Eq. 10), plotted in magenta for $T = 138$°C, describes the permittivity in the SmAP$_F$ well. (C) The nearly linear dependence of $\Delta\varepsilon$, the low-frequency asymptote of $\varepsilon_A'(\omega)$, on cell thickness $d$ is evidence for the PCG mode, although this was not originally recognized as such. The green ellipse highlights the low-frequency ionic current feature in the non-polar SmA phase, which disappears in the SmAP$_F$ phase because of polarization charge screening of the $E$ field in the SmAP$_F$. The interpretation of these dependences of $f_r$ and $\Delta\varepsilon$ on $d$ is identical to that found in the N$_F$ in *Fig. C1*. Adapted from Ref [23]. (Magenta curves and green ellipses added.)

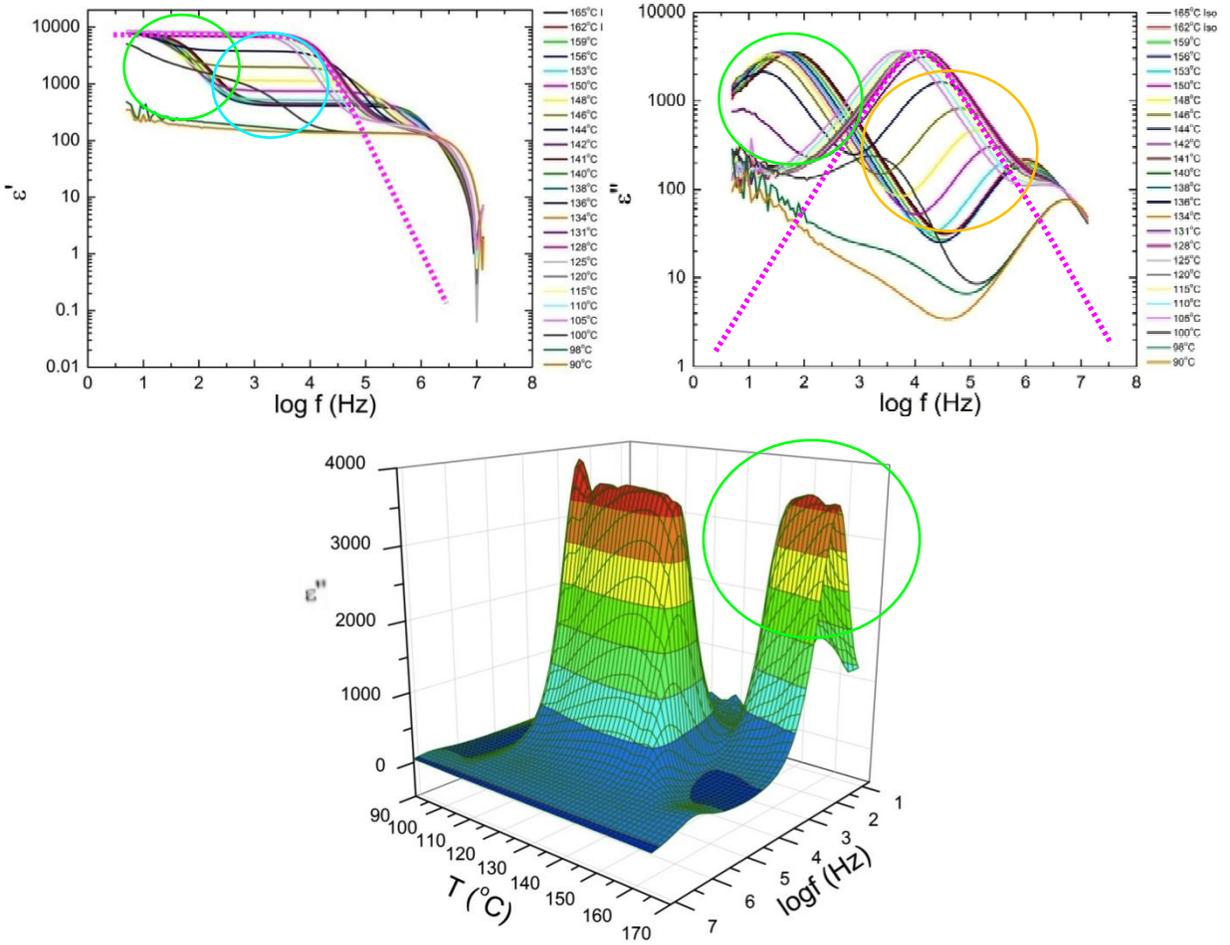

*Figure 8*: Dielectric spectra, reproduced from Ref [43] with permission, showing $\varepsilon_A'(\omega)$ values obtained on cooling a $d = 3.4$ μm cell of the bent-core mesogen W586 into the SmAP$_F$ phase. The polarization in the SmAP$_F$ phase varies from $P = 0.10$ μC/cm$^2$ at $T = 138$°C to $P = 0.35$ μC/cm$^2$ at $T = 105$°C. The smectic layers have bookshelf geometry, with $P$ parallel to the cell plates. The spectra show little dependence on $T$ in the SmAP$_F$ range, overlapping strongly. Specifically, the measured $\varepsilon_A'(\omega \to 0)$ and the peak values of $\varepsilon_A''(\omega = 1/\tau_o) = \varepsilon_A'(0)/2$ are nearly the same for all of the different temperatures. The ion current peak, circled in green, disappears abruptly at the SmA to SmAP$_F$ transition. The scaled PCG model for Re $Q_\omega$ (Eq. 10), overplotted in magenta for $T = 140$°C, describes the major features of $\varepsilon_A(\omega)$ in the SmAP$_F$ phase quite well. The temperature dependence of $\varepsilon_A''(\omega)$ in the cyan-orange circled frequency regimes indicates pretransitional ferroelectric soft-mode behavior. A notable feature of these data is the overlap of the high-frequency falloff of the $\varepsilon_A''(\omega)$ curves, also a tendency evident in *Fig. 4* for the N phase (80/20 and 70/30 mixtures). In a PCG-like model, this overlap would indicate that the effective $R_{LC}$ associated with the soft mode is independent of $T$. Then, since $\tau_o = R_{LC}C_I$, and $\varepsilon_A''(f_o) = \varepsilon_A'(f)/2 \propto C_I$, we would have $\varepsilon_A''(f_o) \propto \tau_o = 1/f_o, \propto C_I$, which is the behavior observed inside the gold circle. Adapted from Ref [43] (green circle added).

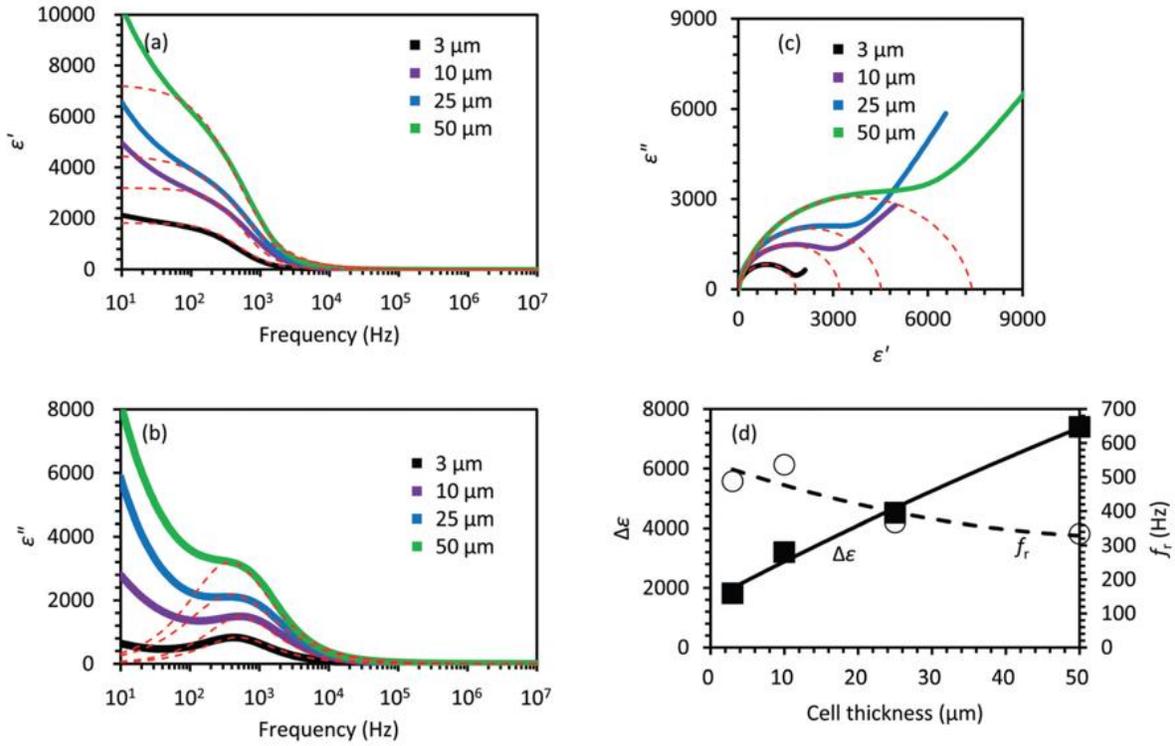

*Figure 9*: (a-c) Dielectric spectra, reproduced from From Ref. [24, open access] with permission, showing $\varepsilon_A'(\omega)$ values obtained from cells of different thickness upon cooling of the bent-core mesogen 4OAM5AMO4 in the SmAP$_F$ phase. The layer geometry is bookshelf, with *P* parallel to the cell plates. The scaled PCG model for Re $Q_\omega$ (Eq. 10) plotted as the red dashed lines, describes the major features of the frequency dependence of the permittivity in the SmAP$_F$ range reasonably well. (d) The approximate linear dependence of $\varepsilon_A'(\omega)$ on cell thickness *d* is evidence for the PCG mode, although this was not originally recognized as such in this paper. The asymptotic low-frequency dependence $\varepsilon_A'(\omega)$ ~1/$\omega$ indicates resistive leakage through the capacitive interface layer.

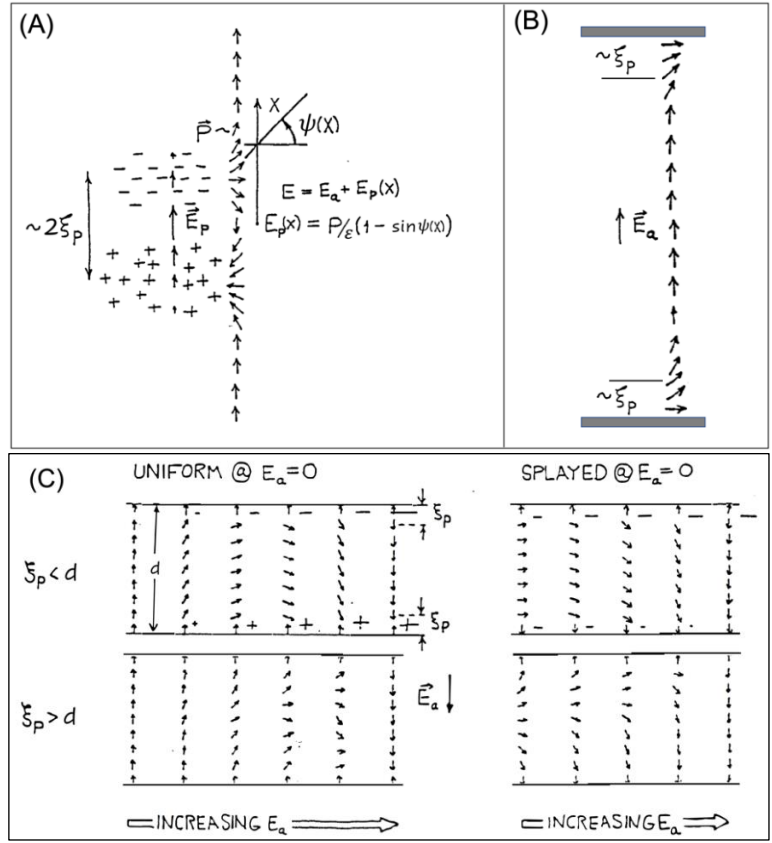

*Figure A1*: (A) One-dimensional, polarization-stabilized kink (PSK) in a 3D space, with $\psi$ starting at $\pi/2$ and reorienting through $2\pi$ along $x$, reproduced from Ref. [26] with permission. In 3D, this $2\pi$ reorientation of $P(x)$ is stabilized by the attraction of the sheets of polarization charge of opposite sign shown in the sketch. The kink in the orientation has a width $\xi_P =$ determined by the balance of electrostatic and Frank elastic torques. In the absence of applied electric field ($E_a = 0$), $P(x)$ is a solution to Eq. 21a,b of the form $\psi(x) = 2\tan^{-1}(x/\xi_P)$. (*B*) PSK solution for the director field in a planar-aligned cell with the boundary conditions with parallel polar orientation on the two surfaces, in an applied field $E_a$. Here $V = V_{sat}$, $\psi(x = \pm d/2) = \pi/2$, and the reorientations at the surfaces have a surface energy/area cost of $\sim K/\xi_P$, which for typical $N_F$ phases is orders of magnitude larger than typical nematic Rapini-Papoular surface anchoring energy/area. Under this condition, these surface kinks are forced out of the LC layer and *P* becomes uniform throughout, with polarization charge confined to the $N_F$ surfaces. (*C*) Director field reorientation in an increasing applied field in sandwich cells of liquid crystal with high and low polarization, shown for the cases where the initial, field-off state is either uniform or splayed. In the high-polarization limit (top row), the polarization in the interior of the cell reorients as a uniform block and any non-uniformities of the polarization field are confined to the cell boundaries. In the low-polarization case (bottom row), the elastic coupling causes the polarization to reorient continuously across the cell as the field is increased, until, at the highest field strengths, the distortions of the polarization field are also confined to the cell surfaces.

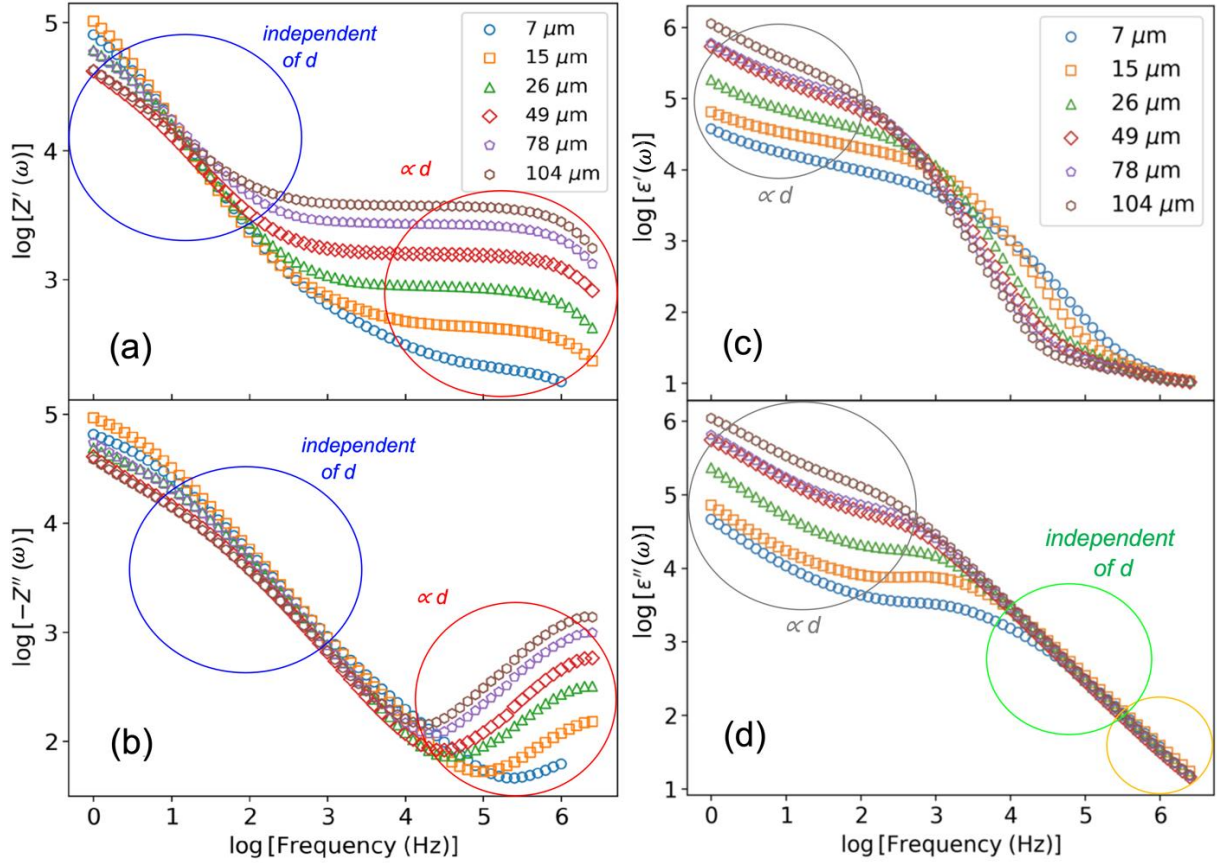

*Figure C1*: (*a,b*) Broadband measurements of the frequency-dependent impedance $Z(\omega)$ of plane-parallel capacitors filled with RM734 [56]. (*c,d*) The resulting apparent frequency-dependent dielectric constant $\varepsilon(\omega)$, calculated using the homogeneous dielectric assumption, Eq. 13a as $\varepsilon(\omega) = (d/A)[1/i\omega Z(\omega)]$. These plots can be understood by taking $Z(\omega)$ to be the sum of interfacial and bulk impedances. At low frequency the interfacial impedance, which is independent of sample thickness $d$ (blue circles), dominates. The result from Eq. 13a is an apparent $\varepsilon(\omega)$ which is proportional to $d$ (gray circles). At high frequency the interfacial capacitive impedance decreases leaving the bulk impedance, which is proportional to $d$ (red circles). The result is a bulk dielectric constant, independent of $d$ (green circle). Adapted from Ref. [56] (circles and associated annotation added).